\documentclass[12pt,a4paper]{article}
\usepackage[utf8]{inputenc}
\usepackage[top=3.2cm,bottom=3.2cm,left=2.5cm,right=2.8cm]{geometry}
\usepackage{comment}
\usepackage{graphicx}
\usepackage{float}
\usepackage{amssymb}
\usepackage{amsmath}
\usepackage{esvect}
\usepackage{mathtools}
\usepackage{subfigure}
\usepackage{bm}
\usepackage{xcolor}
\usepackage[toc,page]{appendix}
\usepackage{parskip}
\usepackage{amsthm}
\usepackage{mathrsfs}
\usepackage{amsfonts}
\usepackage{bbm}
\usepackage{fancyhdr}
\usepackage{booktabs}
\usepackage{authblk}
\usepackage[style=apa]{biblatex}
\numberwithin{equation}{section}
\usepackage[urlcolor=red]{hyperref}
\usepackage{multirow}

\newcommand{\R}{\mathbb{R}}

\providecommand{\keywords}[1]
{
  \small	
  \textbf{Keywords: } #1
}
\usepackage[labelfont=bf]{caption}

\title{Influence of advanced footwear technology on sub-2 hour marathon and other top running performances}
\author[1]{Andreu Arderiu}
\author[2]{Raphael de Fondeville}
\affil[1]{{\footnotesize École Polytechnique Fédérale de Lausanne (EPFL)}}
\affil[2]{\footnotesize Swiss Data Science Center (SDSC)}
\date{\today}

\begin{document}
\maketitle

\begin{abstract}
In $2019$, Eliud Kipchoge ran a sub-two hour marathon wearing Nike's Alphafly shoes. Despite being the fastest marathon time ever recorded, it wasn't officially recognized as race conditions were tightly controlled to maximize his success. Besides, Kipchoge's use of Alphafly shoes was controversial, with some experts claiming that they might have provided an unfair competitive advantage.
In this work, we assess the potential influence of advanced footwear technology and the likelihood of a sub-two hour marathon in official races, by studying the evolution of running top performances from 2001 to 2019 for long distances ranging from $10$km to marathon.
The analysis is performed using extreme value theory, a field of statistics dealing with analysis of rare events.
We find a significant evidence of performance-enhancement effect with a $10\%$ increase of the probability that a new world record for marathon-men discipline is set in $2021$. However, results suggest that achieving a sub-two hour marathon in an official race in $2021$ is still very unlikely, and exceeds $10\%$ probability only by $2025$.  \end{abstract}


\section{Introduction} 
In 2017, Nike officially released the new generation Vaporfly 4\% shoes with slogan ``Designed for record breaking speed''. As a part of its advertisement campaign, the brand initiated the ``Breaking 2'' project, with the aim to break the two-hour marathon barrier.
Since then, the sport community had growing suspicions that the 2017-released shoes, and subsequent models using similar Advanced Footwear Technology (AFT), have a non-negligible effect on running performance, with some voices questioning the fairness of the competitive advantage they might provide.
The technology behind these models includes a very light and responsive foam sole combined with an embedded curved carbon fibre plate, which has been shown to give an energetic advantage to athletes wearing them (\citeauthor{hoogkamer_comparison_2018}, 2018 and  \citeauthor{barnes_randomized_2019}, 2019). Therefore, these controversial shoes sparked a vivid debate leading to new regulations: In January $2020$, World Athletics (\citeauthor{regulations_ban_2020}, 2020) imposed a ban on any shoe with a sole thicker than 40 mm, and limited the maximum number of carbon plates to one. This situation is reminiscent of the $2010$ controversy in elite swimming of Speedo's record-breaking full body swimsuit.

Several studies attempted to quantify the influence of Nike's (AFT) on running performance: \citeauthor*{hoogkamer_comparison_2018} (2018) conducted laboratory experiments with runners and found that Vaporfly's reduced the energetic cost by an average of $4\%$, giving its name to the first model. Later,  \citeauthor{barnes_randomized_2019} (2019), monitored biomechanical and physiological variables to assess the effect of carbon fibre new generation shoes on long distance runners. They confirmed the presence of a $4\%$ energy reduction in average compared to other popular racing shoes. However, the analysis was performed on small groups of sub-elite runners in short trials of 5 min per shoe, limiting their relevance. Besides, note that a $4\%$ metabolic saving would theoretically translate into just a 2-3$\%$ increase in running performance (\citeauthor{kipp2019extrapolating}, 2019). In parallel, Wired Magazine (\cite{thomson_nikes_2017}) performed simple data analysis on running times achieved during the New York City Marathon by amateurs wearing Vaporfly shoes and found that, on average, they ran the second half of the race faster than other participants. Similarly, in a subsequent large-scale statistical analysis, \cite{quealy_nikes_2019} found that Vaporfly users ran from $2\%$ to $5\%$ faster in marathons and half marathons. However, the latter study had limited scientific value as data was extracted from self-reported information on Strava, a fitness app for athletes. Recently, \citeauthor*{guinness_observational_2020} (2020) compared marathon running-times of elite runners, with and without Vaporfly shoes, and estimated a performance increase of $1\%$ to $3\%$.

 All these works just focused on the impact of AFT on average performances, but several recent papers have also evaluated their effect on top performances: \citeauthor{bermon2021effect} (2021) estimated a $0.5\%$ to $2\%$ decrease of top seasonal best times in 10-km races, half-marathons and marathons, from 2016 to 2019, and found that the adoption of this new footwear technology had a significant impact on running times. \citeauthor{senefeld2021technological} (2021) similarly found that the introduction of AFT contributed to a $2\%$ to $3\%$ improvement of best performances during the World Marathon Major series. 
 
 In this work we also analyze fastest running times over the past few years, but we take a different approach, modeling top times as rare extreme events, i.e., large deviations from average running times.
 Extreme value theory (EVT) is a branch of statistics that specifically deals with such extremes and that has been successfully applied to analyse athletics performances in various context: \citeauthor{strand_modeling_1998} (1998) analysed the relation between age and performance for 10km road race athletes, and estimated their age of peak performance.
 \citeauthor{blest_lower_1996} (1996) analysed historical world records for various athletic disciplines to assess the existence of best achievable performances. \citeauthor{robinson_statistics_1995} (1995) analysed women's $1500$ and $3000$m running times to estimate the best achievable performance for women's 3000m track, and assess if a recently broken record was susceptible to be achieved under drug enhancement. Later, \citeauthor{stephenson_determining_2013} (2013), used data from different Olympic disciplines, both for men and women, to compare the history of world records across disciplines. In a different fashion, \citeauthor*{einmahl_records_2008} (2008) and  \citeauthor{rodrigues_statistics_2011} (2011) compared the quality of world records for different disciplines by estimating their best achievable performance.

 In our study, we aim to quantify the influence of AFT on fastest times, i.e., assessing their impact on the frequency that a distance is ran under a given time in a given year, and on the corresponding running times. We propose a statistical model allowing to estimate the probability that a sub-2 hour marathon is run in a given year while accounting for the potential competitive advantage given by AFT.
 In this regard, similarly as \citeauthor{spearing_ranking_2021} (2021) did for elite swimmers data,  we leverage extreme value theory to compare the effect of new generation shoes across sexes and distances, while accounting for the improvement over time of running techniques and training practices. 
 
 The paper proceeds as follows: Section 2 gives a detailed description of the data, as well as the methodology applied in our study. In Section 3, we present our main conclusions, including expected next records, the likelihood of a sub-two hour marathon, the probability of breaking world records, and running-times adjusted to correct for the AFT effect. We report a significant evidence of AFT performance-enhancement effect, accounting for a $10\%$  increase  of  the  probability  that  a  new  world  record  for  marathon-men discipline is set in 2021.  However, results suggest that achieving a sub-two hour marathon in an official event in 2021 is still very unlikely, and exceeds $10\%$ probability only by 2025. Finally, Section 4 concludes by discussing some limitations of our model, and suggests directions for further improvements.
 
 \section{Methodolgy}
\subsection{Data Exploration}\label{sec:explo}
In this study, we extract yearly top-100 running-times, in seconds, over the period $2001$ to $2019$ for marathon, half-marathon and 10km road disciplines; only official events as labelled by World Athletics are considered. For each athlete, we keep only his, or her, best performance per distance and per year. As it is commonly done in sports data analysis, we consider men's and women's data as different disciplines yielding a total of six disciplines with 1900 data points each. Even though since 2011 World Athletics (\citeauthor{IAAF_2011}, 2011) considers “women only” and "mixed gender" road races as different categories, we cannot afford such distinction as it would reduce the number of observations to a critically low sample size, so questioning the relevance of our study. Also, the presence of pacemakers in men races has a non-negligible impact on the performance, but no data is available to discriminate between "paced" and "unassisted" records, so we choose for consistency to make such distinction neither for women nor men.

Our aim is to estimate the probability that a running time drops below a given reference.
In this setting, best performances correspond to the shortest race times, and we focus on extremely short running times, that can be viewed as large negative deviations from the mean.
A natural tool to analyse such extreme events is Extreme Value Theory, and in particular Peaks-Over-Threshold (POT) analysis: the methodology provides a framework to approximate the distribution of exceedances, i.e., the probability, and its frequency, that a variable drops below a given threshold.
In practice, we simply fit a statistical model to any data point that exceeds a large negative threshold.

The mathematical formulation of the model is theoretically justified as it corresponds to the universal approximator of the distribution of independent exceedances.
For this reason, the model can be used for extrapolation, i.e., quantify the probability of running times that have not been observed yet.
An important parameter, the tail index, determines the regime of extrapolation: a positive tail index implies that any running time below the threshold has a positive probability of occurrence; while for a negative tail indexes observations are lower bounded.
In sports, multiple studies, e.g., \citeauthor{robinson_statistics_1995}, 1995; \citeauthor{blest_lower_1996}, 1996; \citeauthor{strand_modeling_1998}, 1998;  \citeauthor{einmahl_records_2008}, 2008; \citeauthor{rodrigues_statistics_2011}, 2011, found negative indexes giving strong evidences in favor of the existence of a best achievable performance.
Similar analysis have also been performed in other context such as life expectancy (\citeauthor{einmahl_limits_2019}, 2019), natural hazards (\citeauthor{holmes_statistical_2008}, 2008) or hydrology (\citeauthor{katz_statistics_2002}, 2002). Throughout this article, we assume independence between running-times across distances and years. In practice, this means that an athlete can contribute multiple times to the data set, but only once per year and per distance. Such hypothesis is common, and necessary, for the analysis of time dependent data using extreme value theory; see \citeauthor{spearing_ranking_2021} (2021) for a thorough discussion about the impact of possible residual dependence.

\begin{figure}[H]
\centering
    \subfigure{\includegraphics[width=0.45\textwidth]{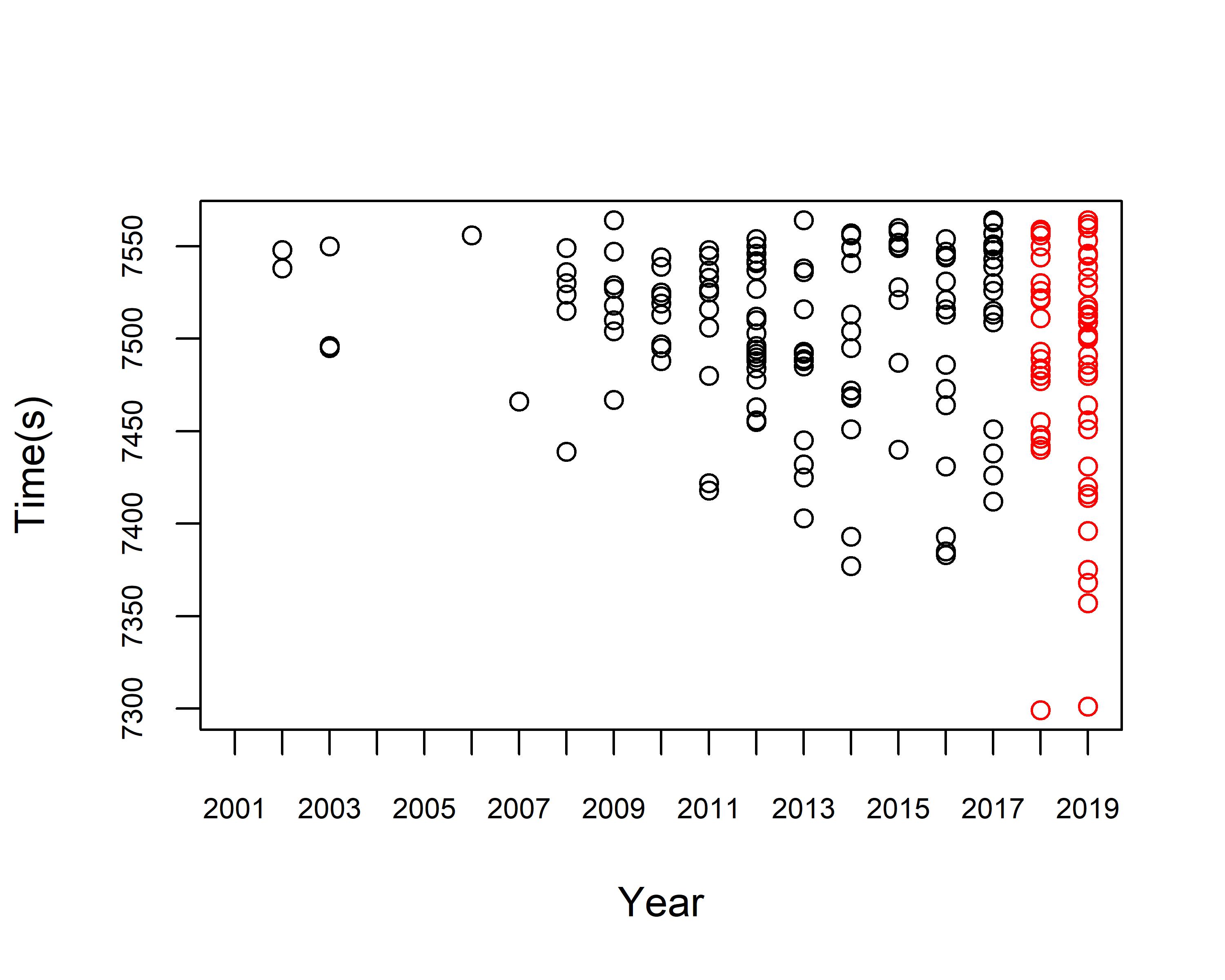}}\hspace{0.3cm}
    \subfigure{\includegraphics[width=0.45\textwidth]{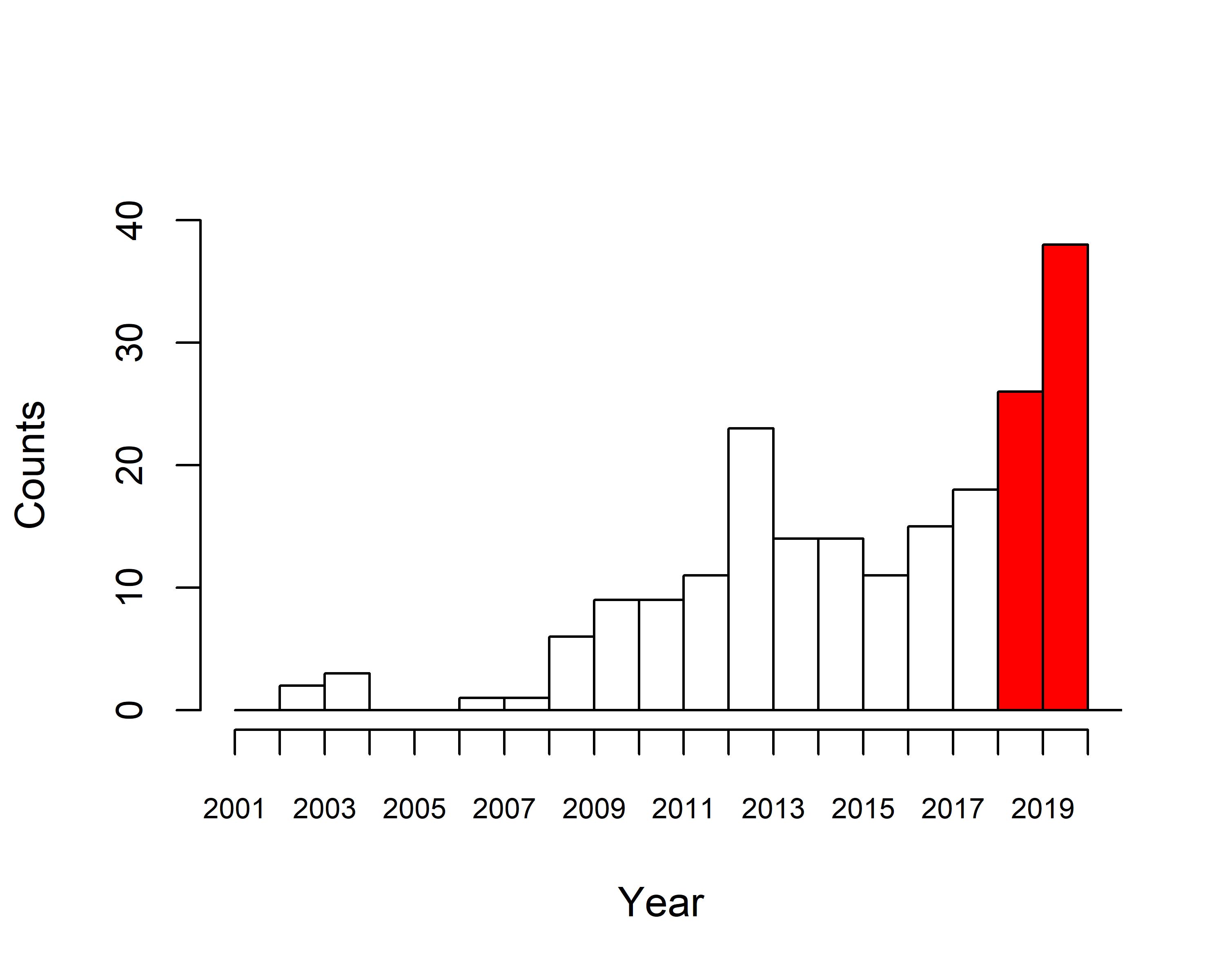}}
   \\
    \caption{Men's marathon best running times from $2001$ to $2019$. Left: $200$ best all-time marathon running-times per year. Right: yearly number of running times within a given year. Red points (bars) correspond to observations (years) for which shoes with technological advances are widely used in official races.}
    \label{1}
\end{figure}

 For each discipline, we select
 a threshold such that over the period $2001$ to $2019$ there are exactly $200$ running-times that drop below; data for the men's marathon is displayed in Figure \ref{1}. Threshold selection was performed using mean residual life plots (\citeauthor{coles_introduction_2001}, 2001), which can be found in Appendix \ref{app: plots}. We observe a temporal increase in both race time performance, and frequency at which exceedances occur. We also note a noticeable step increase in year $2018$ and $2019$, corresponding to the democratization of AFT amongst elite runners (\cite{quealy_nikes_2019}). The unexpected and sudden frequency increase in $2012$, which is also present in women marathons, has no clear explanation; \citeauthor{bermon2021effect} (2021) hypothesized that the relatively cooler global temperatures in 2012 (\citeauthor{weather_NASA}, 2020) might have boosted endurance performance and facilitated better marathon results that year. Similar trends are observed across all disciplines. 

\subsection{Model}\label{sec:model}
For each discipline, we model running times dropping below their respective thresholds following the work of \citeauthor{spearing_ranking_2021} (2021) for elite swimmers; technical methodological details can be found in Appendix \ref{app:model}. The model provides estimates for the expected number of exceedances per year, as well as the probability that the running time falls below given lower references. Furthermore, as we find negative tail indexes for all disciplines, the model provides an estimate for the shortest achievable running time within a year that we call ultimate time.

The model includes time-dependent parameters to account for the improvements of racing and athletes conditions over time, as well as for a ''AFT effect" that we assume to appear in $2018$. We acknowledge that in 2017 some athletes were already competing with Vaporfly shoes, but we found no statistically significant AFT effect in 2017. Indeed, according to the analysis by \citeauthor{thomson_nikes_2017} (2017) and \citeauthor{bermon2021effect} (2021), their adoption was still very limited, and it was not until 2018 and 2019 that they started to be widely used by runners. Multiple models with different temporal dependencies were considered, but none of them were significantly better than the presented model. Parameter estimates retrieve for all disciplines a positive temporal trend for improvements in running techniques and training practices, which is specially significant for marathon-men and half marathon-women. Similarly, the AFT effect is significant and positive across all disciplines, with the exception of half marathon-man, indicating an overall benefit of using these advanced technology shoes. There is no clear explanation to the apparent absence of such effect in half-marathon man, which is also reported in \citeauthor{bermon2021effect} (2021).  Finally, the AFT impact is stronger for women than for men: for instance, the effect is $200\%$ stronger for marathon-women than for marathon-men. Such unequal impact across female and male runners is also found by \citeauthor{bermon2021effect} (2021) and \citeauthor{senefeld2021technological} (2021).

To assess the overall quality of the fitted model, we compared yearly frequencies and running times faster than their respective threshold, to the theoretical quantities provided by the fitted model and found an overall good fit; see Appendix \ref{app: plots}.

\section{Results}
\subsection{Yearly ultimate times}
We computed the ultimate times for all six disciplines as function of time: these change linearly with time accounting for continuous improvement of techniques and preparation over time. Table \ref{tb:ratios} displays the world records as in $2019$ for all disciplines against estimated ultimate times for $2019$ and $2025$: we observe a substantial decrease for marathon-men from $2019$ to $2025$. In contrast, the ultimate time for marathon-women just decreases few seconds over the same period. Indeed, during the last two decades, changes in men's marathon world leading times have been much faster than in women's marathon, so this directly reflects on the evolution of yearly ultimate times.

\begin{table}[H]
\centering
\caption{World records as in $2019$ and ultimate times for $2019$ and $2025$, for all disciplines, with $95\%$ confidence intervals.}\vspace{0.2cm}
\begin{tabular}{c|cll}
discipline             & World record 2019 & Ultimate 2019  & Ultimate 2025  \\ \midrule
Marathon-men      & 02:01:39       & 01:59:43 (-30s,+14s) &  01:58:13 (-33s,+16s) \\
Marathon-women    & 02:14:04        & 02:13:03 (-45s,+12s) & 02:12:55 (-49s,+12s)\\
Half marathon-men & 00:58:01        & 00:57:28 (-8s,+6s) & 00:57:08 (-9s,+6s) \\
Half marathon-women & 01:04:51        & 01:02:39 (-21s,+11s) & 01:01:38 (-19s,+19s) \\
10km-men          & 00:26:38       & 00:26:22 (-6s,+4s)  & 00:26:15 (-6s,+4s) \\
10km-women        & 00:29:43        & 00:29:05 (-10s,+6s)& 00:28:49 (-11s,+7s) \\
\end{tabular}
\label{tb:ratios}
\end{table}

\subsection{Expected next record}
\begin{figure}[H]
\centering
    \includegraphics[width=0.7\textwidth]{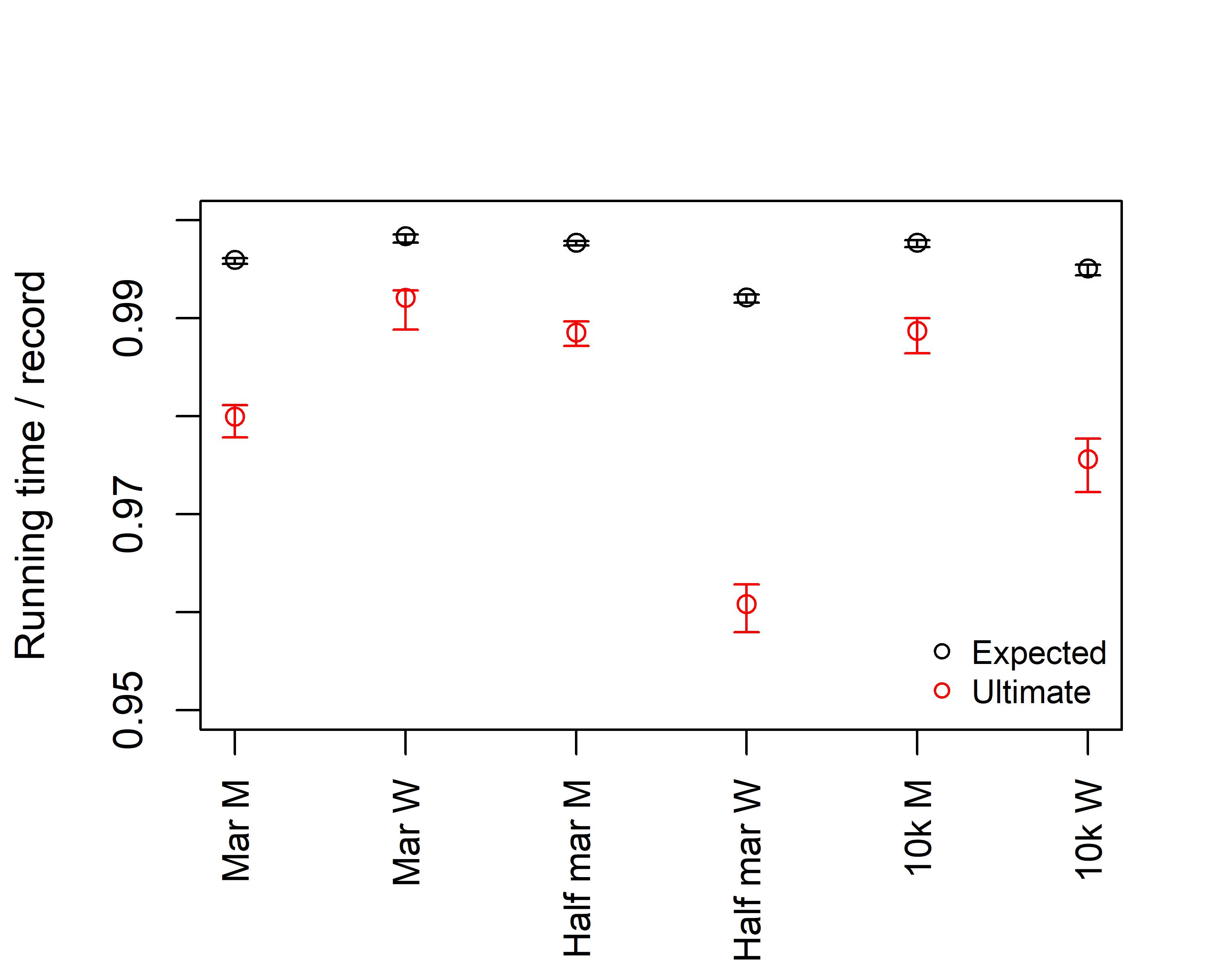}
        \caption{Expected (black) new world records, if it were to be broken in $2021$, and corresponding ultimate times (red) for every discipline, with $95\%$ confidence intervals. For clarity, estimated times are normalised by the current record as of $2019$. }.
    \label{fig:ratios}
\end{figure}

The model provides the probability that a running time drops below given reference times.
We can thus set the reference to the current world record, and estimate the expected running-time of a new record in a given year. Figure \ref{fig:ratios} displays the estimated expected running-time of the next world record for the year 2021 with corresponding ultimate times. Different disciplines have different scales of time, so, to ensure a proper comparison between disciplines, we scale all values by their respective $2019$ world records. As an example, the marathon-men $2019$ world record is $2$ hours $1$ minute $39$ seconds, and the $2021$ ultimate time is $1$ hour $59$ minutes $43$ seconds, so their ratio in seconds is $0.98$. The difference between expected new world record and ultimate times gives an idea of how close we expect the new record to be to the fastest possible time in $2021$. 

In Figure \ref{fig:ratios}, differences between expected new record and ultimate running time vary across different disciplines, ranging from  $1$ minute 1 second, i.e., $0.8\%$, for marathon-women, to $2$ minutes $14$ seconds, i.e., $4.3\%$, for half marathon-women. Expected improvements of world records in $2021$ are slightly smaller in percentage for disciplines where the current record is closer to the fastest possible time, but differences are relatively small ranging from $0.2\%$ for marathon-women to $0.8\%$ for half marathon-women, which stems from the common tail index shared across disciplines. As the ultimate time decreases over time, if the record is not broken then the gap between current record and ultimate time increases, giving range for greater improvement.

\subsection{Probability of record breaking in a given year}
\begin{figure}[H]
\centering
    \includegraphics[width=0.7\textwidth]{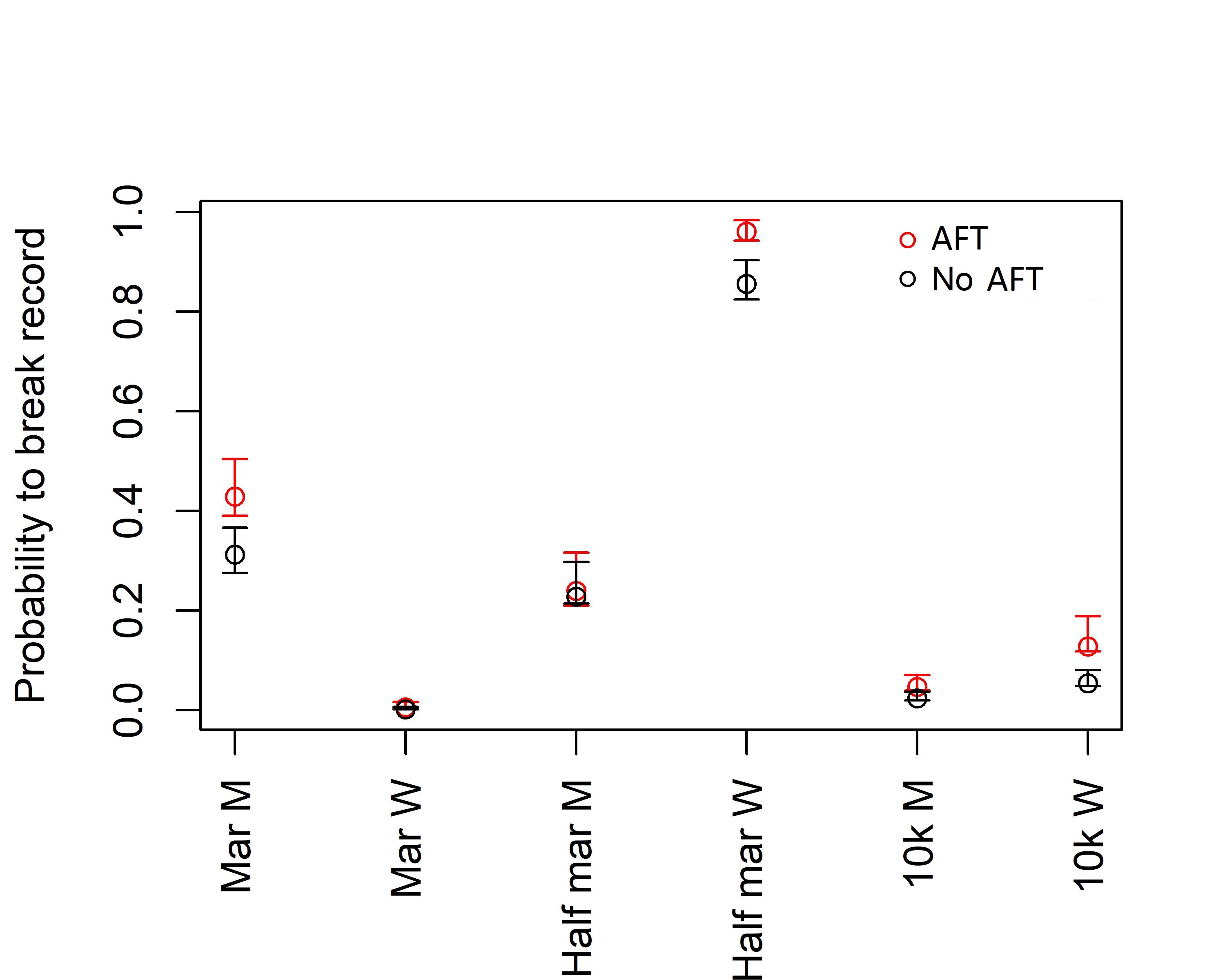}
    \caption{Probability of the world record being broken in $2021$ for all disciplines, with $95\%$ confidence intervals. Red dots: estimates with AFT effect. Black dots: corrected estimates removing the effect of AFT.}.
    \label{fig:brecord}
\end{figure}

We can use the fitted model to estimate the probabilities of breaking a world record in any given year after 2019. Figure \ref{fig:brecord} displays the estimated probabilities of breaking the world record in $2021$ with and without correcting for the effect of AFT.

 We observe how probabilities vary significantly from discipline to discipline, ranging from a $1\%$ chance for marathon-women to a $96\%$ chance for half marathon-women. Such low chance for marathon-women is coherent with the fact that the difference between its current world record and $2021$ ultimate time is the smallest across disciplines, so its record might be harder to break than for other disciplines. Estimates correcting for the effect of AFT are extremely similar for marathon-women, half marathon-men, and 10km-men, which contrasts with the substantial probability drop of about $10\%$ for marathon-men and half marathon-women, and of $8\%$ for 10km-women.

\subsection{Time until next record breaking}
In the previous section we estimated the probability of breaking a record in a given year. In a similar fashion, we can use the fitted model to estimate the probability of the current record to be broken before a given year.
These are computed for consecutive years, and we find for each discipline the earliest year for which such probability exceeds $95\%$. In other words, we estimate the expected waiting time to observe a new record, with at least $95\%$ certainty;
results are displayed in Table \ref{95} with and without AFT adjustment.

For marathon-men the world record will most likely be broken before $2024$, and after correcting for the AFT effect, the year estimate increases by just one. In contrast, for disciplines with greater estimated year, as for marathon-women, the impact of AFT effect is much more substantial. Similarly to Figure \ref{fig:ratios}, comparison between disciplines can be summarized by the ratio of estimated waiting times with and without AFT effect: these ratios lie between $0.70$ and $0.85$ for all disciplines, except half-marathon men, where the AFT effect is not significant as explained in Section \ref{sec:model}.



\begin{table}[H]
\centering
\caption{Estimated earliest year before which there are $95\%$ chances that the current world record is broken, with $95\%$ confidence intervals. AFT years correspond to estimates with the AFT effect, while the rightmost column correspond to the estimates corrected to remove the influence of the AFT effect.}\vspace{0.2cm}
\begin{tabular}{c|cc}
discipline             & Year AFT &  Year AFT-corrected\\ \midrule
Marathon-men      & 2024 (-0y,+0y)          & 2025 (-1y,+0y)             \\
Marathon-women    & 2042 (-6y,+0y)          & 2048 (-4y,+0y)             \\
Half marathon-men & 2027 (-1y,+1y)          & 2027 (-1y,+1y)             \\
Half marathon-women & 2022 (-1y,+0y)          & 2022 (-0y,+0y)           \\
10km-men          & 2036 (-1y,+1y)         & 2040 (-3y,+0y)             \\
10km-women        & 2029 (-1y,+1y)          & 2033 (-2y,+0y)          \\  
\end{tabular}
\label{95}
\end{table}

Similarly, we can further estimate the expected waiting time until the current world record is broken for each discipline. Table \ref{tb_exp} shows 
the estimates for the expected waiting times, and their corrections obtained by removing the AFT effect. We observe for marathon-men that the current world record is expected to be broken in $2$ years, which contrasts with marathon-women, where the expected waiting time is $15$ years. For the rest of disciplines, it can be seen that expected waiting times are below $10$ years. It is also remarkable that when neglecting the AFT effect, waiting times substantially increase for all disciplines but marathon-men, half marathon-men and half marathon-women. This is coherent with the previous analysis made for Table \ref{95}.

\begin{table}[H]
\centering
\caption{Expected waiting time, in years with $95\%$ confidence intervals, until next record is set for all disciplines. AFT times correspond to estimates with the AFT effect, whereas the rightmost column correspond to times corrected to remove the AFT effect.}\vspace{0.2cm}
\begin{tabular}{c|cc}
discipline             & Time AFT    & Time AFT-corrected \\ \midrule
Marathon-men      & 2.2 (2.1,2.3)    & 2.5 (2.2,2.6)    \\
Marathon-women    & 15.4 (11.6,15.6) & 20.6 (16.3,21.0) \\
Half marathon-men & 3.7 (3.2,4.1)    & 3.8 (3.3,4.1)    \\
Half marathon-women & 1.1 (1.1,1.1)    & 1.3 (1.2,1.3)    \\
10km-men          & 8.7 (7.1,9.3)    & 11.3 (9.4,12.0)  \\
10km-women        & 4.3 (3.5,4.7)    & 6.5 (5.5,6.9)   \\
\end{tabular}
\label{tb_exp}
\end{table}

\subsection{Corrected times without the AFT effect}
In a similar fashion as \citeauthor{spearing_ranking_2021} (2021) did for the use of full body suits in swimming, we can adjust running-times for the use of the shoes with AFT. More precisely, for a given discipline, the corrected running-time of a performance achieved after the democratisation of AFT in 2018, is computed by matching probabilities of exceedances with and without the AFT effect. As an example, the current world record for 10km-women is $29$ minutes $38$ seconds, which was set in 2021 by Kalkidan Gezahegne wearing shoes with AFT. If we adjust such record time for the AFT effect, we obtain $29$ minutes and $44$ seconds, which represents a correction of $+6$ seconds. This suggests that if modern shoes with AFT hadn't been used in $2021$, the world best would still be the $29$ minutes $43$ seconds, set by Joyciline Jepkosgei in $2017$. Corrections for world records of all other disciplines can be found in Table \ref{tb:corrected} of Appendix \ref{app:model_estimates}.


\subsection{Likelihood of a sub-two hour marathon}
Widespread popular belief claimed that the two-hour marathon barrier was unbreakable. In $2017$ within the project \textit{Breaking 2}, Nike organised a race during which Eliud Kipchoge set a time of $2$ hours and $25$ seconds. In $2019$, Ineos organised the $1$:$59$ Challenge race, where Kipchoge successfully broke the barrier, achieving a time of $1$ hour $59$ minutes $40$ seconds. However, neither of those records are officially recognised, as race conditions were controlled and a rotating cast of pacers shielded Kipchoge from wind throughout the run. Indeed, in Table \ref{tb:ratios} the estimate for the ultimate or fastest possible running-time of marathon-men in $2019$ is of $1$ hour $59$ minutes $43$ seconds, which suggests that even though a sub-two hour marathon would have been theoretically possible, the time achieved by Kipchoge in the Ineos challenge would have been very unlikely with regular official race conditions.

Some studies attempted to predict the year when the first sub two-hour marathon would be achieved: \citeauthor{joyner_two-hour_2011} (2011) estimated the rate of improvement of marathon-men world records since the late $1920$s, finding that a time under $2$h could occur between $2021$ and $2036$. \citeauthor{angus_statistical_2019} (2019) used marathon world record performance times since $1950$, and estimated that the probability of observing a sub-two hour marathon in $2020$ is just about $3\%$, with chances increasing to $10\%$ by $2032$. However, most of these studies analyze data sets that are not representative of athletics current state as they do not reflect recent changes in sport practices and usually suffer from selection bias.

The estimated probability of a sub-two hour marathon in $2020$ obtained with our model is of $0.1\%$ $(0.04\%,0.3\%)$, much lower than the estimate provided by \citeauthor{angus_statistical_2019}. Such discrepancy can be explained by the fact that, while we base our analysis on 200 top times for each discipline, \citeauthor{angus_statistical_2019} just use world record progression data with a total of $26$ data points, so their estimates might suffer from high variability. Still, both results agree that it is still very unlikely that without controlling race conditions or offering additional support for runners a sub two-hour marathon can be achieved in $2020$.

\begin{figure}[H]
\centering
    \subfigure{\includegraphics[width=0.48\textwidth]{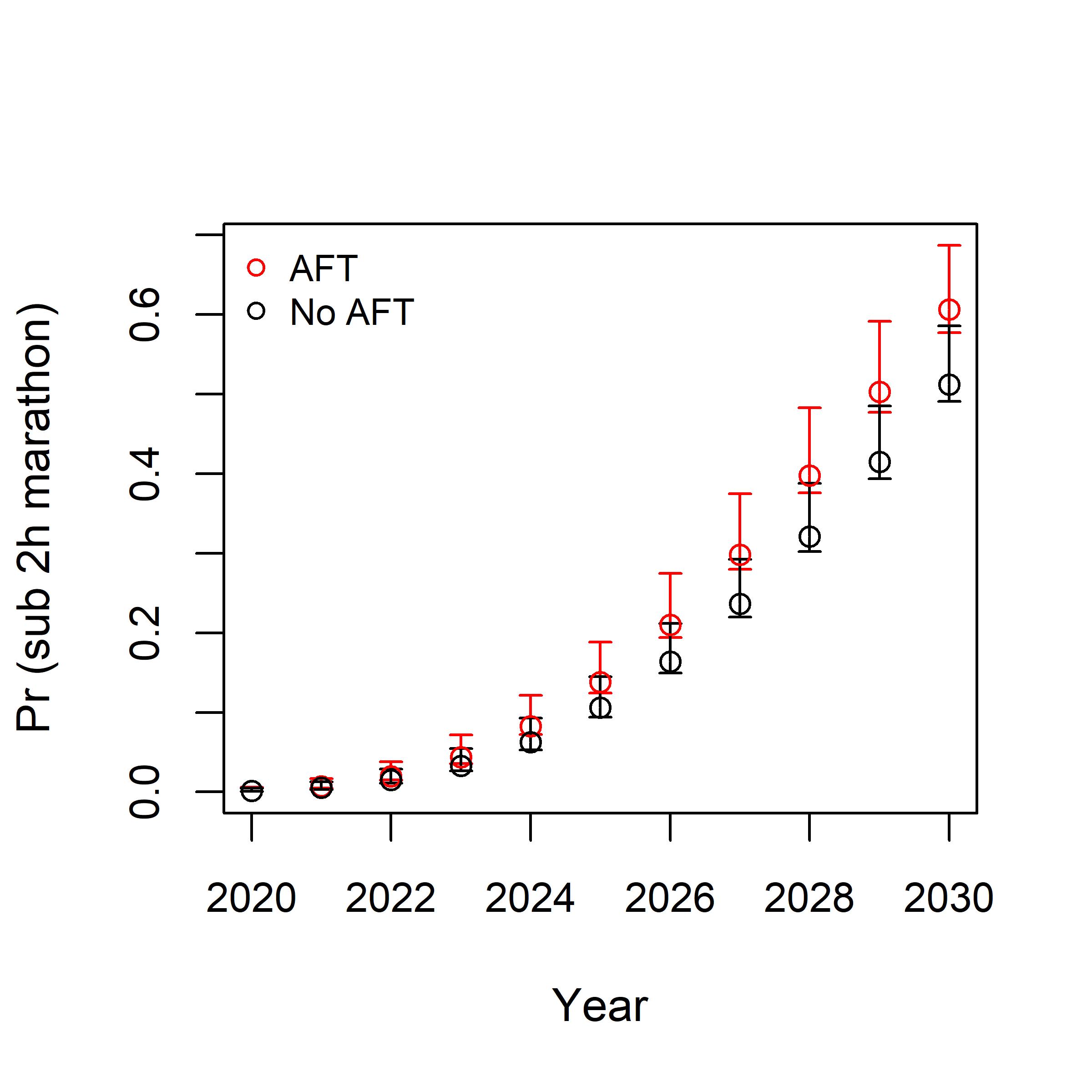}}
    \subfigure{\includegraphics[width=0.48\textwidth]{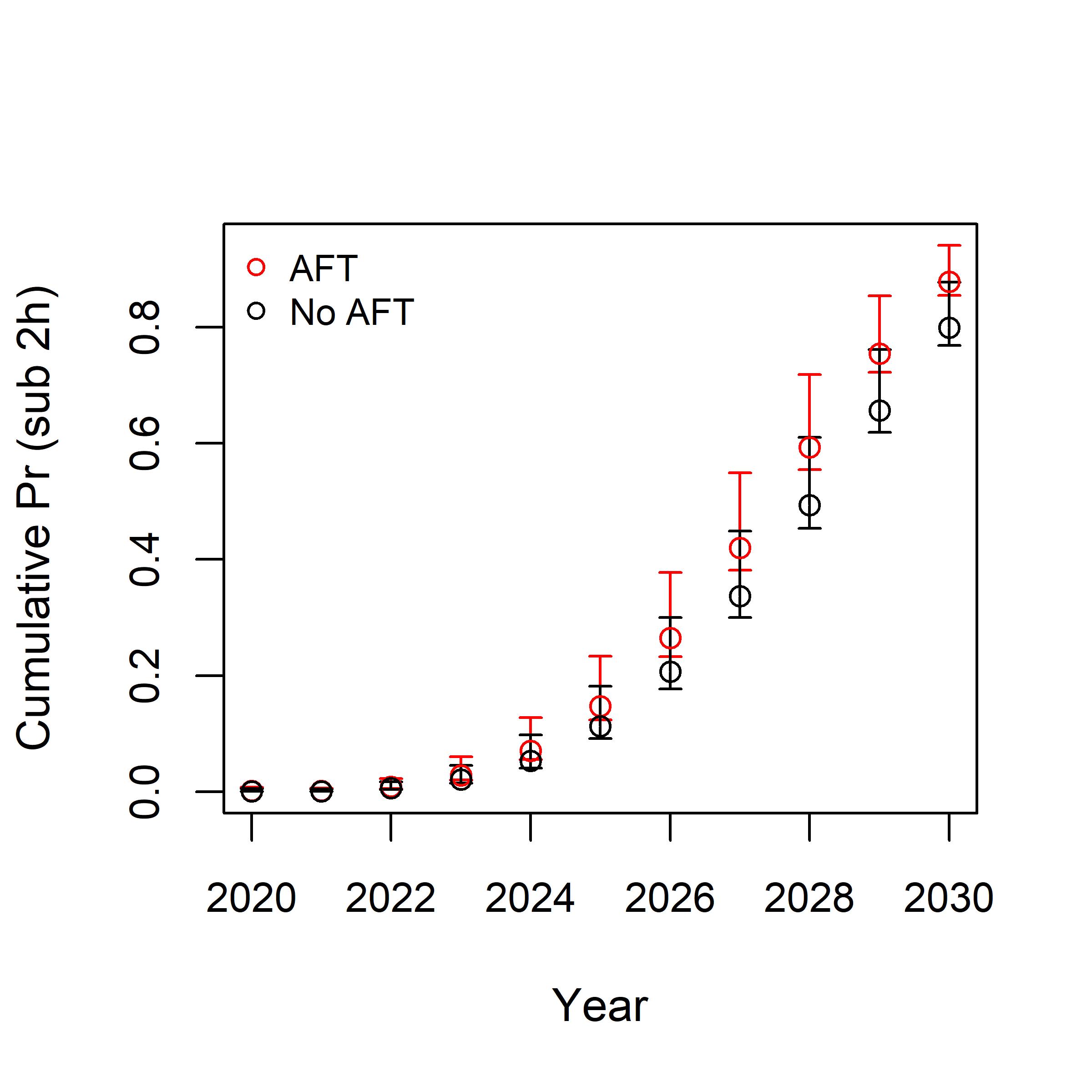}}
    
    \caption{Probability that the a sub two-hour marathon is achieved by a man in a given year (left), and before a given year (right), for the $2020$-$2030$ period with $95\%$ confidence intervals. Red dots correspond to the probability computed with advanced technology shoes, whereas black dots are corrected for such AFT effect .}
    \label{fig:sub2}
\end{figure}

Additionally, we compute estimates for the probability that a sub two-hour marathon is achieved in a given year. Figure \ref{fig:sub2} (left) displays such estimates for the $2020$-$2030$ period, with and without the AFT. We observe how before $2025$ all probabilities are below $10\%$, and the chances of breaking the two hour barrier with and without the AFT effect aren't significantly different. Note also that $2030$ is the first year where the chances of breaking the barrier exceeds $50\%$. In that case, if we neglect shoes effect, chances fall to around $40\%$. Figure \ref{fig:sub2} (right) displays cumulative probability estimates, so the chances that a sub two-hour marathon is achieved before a given year.
We observe that there are about $10\%$ and $50\%$ chances that a sub two-hour marathon is achieved before year 2025 and year 2028, respectively. 
Finally, the expected sub two-hour marathon arrival time is found to be 2027, which is coherent with the 2021-2035 range estimated by \citeauthor{joyner_two-hour_2011} in 2011.\\

We analysed in detail the 2-hour marathon as it has been a symbolic threshold popularized by recent events, but it is relevant only for men. So, it is natural to wonder what would be an equivalent feat for women. In 2015 \citeauthor{hunter2015two} analysed top 100 marathon men and women times since 1960 and estimated that the $2$ hour $15$ minutes $25$ seconds time set by Paula Radcliffe in 2003 was already equivalent to a sub-2 hour mean threshold. With a similar approach, \citeauthor{tucker2017unlikeliness} (2017) analysed top times across a wide range of running disciplines, finding that the actual gender gap was overestimated in \citeauthor{hunter2015two} (2015), and that the record by Paula Radcliffe was yet not equivalent to a sub-2 hour marathon. The most recent study by \citeauthor{angus_statistical_2019} (2019), estimated a much lower landmark at $2$ hour $7$ minutes and $33$ seconds. According to our model, the marathon-women equivalent time of a sub-2 hour men marathon in 2020 is $2$ hours $12$ minutes $56$ second and increases by $+11$ seconds when correcting for the AFT effect. Both estimates lie within the range of estimations made by the above mentioned studies.

\section{Discussion}
The main purpose of this study was to analyse the evolution of the frequency and distribution of top running times from various running disciplines, and assess the possible influence of wearing shoes with technological advancements. We observed how performance substantially improved over time for all disciplines, probably due to the improvements in running techniques and training practices. We also found a significant increase of performance since 2018, which we attributed to the democratization of advanced footwear technology. Such influence is found statistically significant, with greater impact for women than for men, and in some cases it has likely given a competitive edge to establish new world records, e.g., 10km-women. Moreover, our results showed that it is still very unlikely that a sub-two hour marathon is achieved in an official race during the next few years, and that the record achieved by Kipchoge in Ineos Challenge would have been very unlikely without all the additional support and controlled racing conditions.

The model has a good overall fit, and provides a good agreement with historical records. However, we couldn't fully explain the variability of some model parameters, such as the linear trends, across different disciplines. One of the underlying assumptions of our model is that the number of official races held for every discipline doesn't substantially change, so observations for every year are equally weighted. Hence, we might be over-, or under-, weighting observations from years where more, or less, races were held; the year $2020$, excluded of this analysis, would be an obvious example. Such yearly data imbalance could be taken into account for more accurate estimation and forecasting. Furthermore, we didn't account for the different race conditions of the venues, which certainly have an impact in the distribution of times. In this aspect, the model could be improved by adding an additional parameter for each venue to capture their influence in running-times. Finally, we assumed that before $2018$ there were no times recorded with Vaporfly shoes, and after $2018$ all times were set with shoes with technological advancements. To improve our model estimation of the influence of advanced footwear technology, it could be relevant to label each data points as performed with or without these type of shoes, similarly as in \citeauthor{guinness_observational_2020} (2020). 

\section*{Acknowledgement}
We acknowledge Harry Spearing for sharing code that provided good inspiration to our work. The authors received no specific funding for this work.



\printbibliography

@article{davison_models_1990,
	title = {Models for Exceedances over High Thresholds},
	journaltitle = {Journal of the Royal Statistical Society. Series B (Methodological)},
	volume = {52},
	number = {3},
    pages = {393--425},
	author = {Davison, A. C. and Smith, R. L.},
	date = {1990},
}

@article{hoogkamer_comparison_2018,
	title = {A Comparison of the Energetic Cost of Running in Marathon Racing Shoes},
	volume = {48},
	issn = {1179-2035},
	pages = {1009--1019},
	number = {4},
	journaltitle = {Sports Medicine},
	author = {Hoogkamer, Wouter and Kipp, Shalaya and Frank, Jesse H. and Farina, Emily M. and Luo, Geng and Kram, Rodger},
	date = {2018},
	pmid = {29143929},
	pmcid = {PMC5856879},
}

@article{strand_modeling_1998,
	title = {Modeling Road Racing Times of Competitive Recreational Runners Using Extreme Value Theory},
	volume = {52},
	issn = {0003-1305},
	pages = {205--210},
	number = {3},
	journaltitle = {The American Statistician},
	author = {Strand, Matthew and Boes, Duane},
	date = {1998},
}

@article{blest_lower_1996,
	title = {Lower Bounds for Athletic Performance},
	volume = {45},
	issn = {0039-0526},
	pages = {243--253},
	number = {2},
	journaltitle = {Journal of the Royal Statistical Society. Series D (The Statistician)},
	author = {Blest, David C.},
	date = {1996},
}

@article{robinson_statistics_1995,
	title = {Statistics for Exceptional Athletics Records},
	volume = {44},
	issn = {0035-9254},
	pages = {499--511},
	number = {4},
	journaltitle = {Journal of the Royal Statistical Society. Series C (Applied Statistics)},
	author = {Robinson, Michael E. and Tawn, Jonathan A.},
	date = {1995},
}

@article{einmahl_records_2008,
	title = {Records in Athletics Through Extreme-Value Theory},
	volume = {103},
	issn = {0162-1459},
	pages = {1382--1391},
	number = {484},
	journaltitle = {Journal of the American Statistical Association},
	author = {Einmahl, John H. J. and Magnus, Jan R.},
	date = {2008-12-01},
}

@article{angus_statistical_2019,
	title = {A Statistical Timetable for the Sub–2-Hour Marathon},
	volume = {51},
	issn = {0195-9131},
	pages = {1460--1466},
	number = {7},
	journaltitle = {Medicine \& Science in Sports \& Exercise},
	author = {Angus, Simon D.},
	urldate = {2020-06-08},
	date = {2019-07},
}

@article{rodrigues_statistics_2011,
	title = {Statistics of extremes in athletics},
	volume = {9},
	number = {2},
	pages = {127--153},
	journaltitle = {Revstat Statistical Journal},
	author = {Rodrigues, Lígia and Gomes, Maria and Pestana, Dinis},
	date = {2011-06-01},
}

@article{scarrott_review_2012,
	title = {A review of extreme value threshold estimation and uncertainty quantification},
	volume = {10},
	number = {1},
	pages = {33--60},
	journaltitle = {Revstat Statistical Journal},
	author = {Scarrott, C. and {MacDonald}, Anna},
	date = {2012-03-01},
}

@article{stephenson_determining_2013,
	title = {Determining the Best Track Performances of All Time Using a Conceptual Population Model for Athletics Records},
	volume = {9},
	number = {1},
	pages = {67--76},
	journaltitle = {Journal of Quantitative Analysis in Sports},
	author = {Stephenson, Alec and Tawn, Jonathan},
	date = {2013-01-30},
}

@online{thomson_nikes_2017,
	title = {Do Nike's Zoom Vaporfly 4\% Marathon Shoes Actually Make You Run Faster? {\textbar} {WIRED}},
	url = {https://www.wired.com/story/do-nike-zoom-vaporfly-make-you-run-faster/},
	author = {Thomson, Nicholas},
	date = {2017-07-11},
}

@online{IAAF_2011,
	title = {IAAF to continue to recognise existing women’s road-running records},
	url = {https://www.worldathletics.org/news/undefined/iaaf-to-continue-to-recognise-existing-womens},
	author = {IAAF},
	date = {2017-10-09},
}

@online{regulations_ban_2020,
	title = {TECHNICAL RULES},
	url = {https://hmg-prod.s3.amazonaws.com/files/c2-1-technical-rules-amended-on-31-january-2020-1580483189.pdf},
	author = {{World Athletics}},
	date = {2020-01-31},
}

@online{weather_NASA,
	title = {Goddard Institute for Space Studies},
	url = {https://climate.nasa.gov/vital-signs/global-temperature/},
	author = {{NASA}},
	date = {2020},
}

@book{coles_introduction_2001,
	location = {London ; New York},
	title = {An Introduction to Statistical Modeling of Extreme Values},
	isbn = {978-1-85233-459-8},
	pagetotal = {209},
	publisher = {Springer},
	author = {Coles, Stuart},
	date = {2001},
}

@article{barnes_randomized_2019,
	title = {A Randomized Crossover Study Investigating the Running Economy of Highly-Trained Male and Female Distance Runners in Marathon Racing Shoes versus Track Spikes},
	volume = {49},
	issn = {1179-2035},
	pages = {331--342},
	number = {2},
	journaltitle = {Sports Medicine},
	author = {Barnes, Kyle R. and Kilding, Andrew E.},
	date = {2019-02},
}

@online{quealy_nikes_2019,
	title = {Nike’s Fastest Shoes May Give Runners an Even Bigger Advantage Than We Thought},
	issn = {0362-4331},
	url = {https://www.nytimes.com/interactive/2019/12/13/upshot/nike-vaporfly-next-percent-shoe-estimates.html},
	journaltitle = {The New York Times},
	author = {Quealy, Kevin and Katz, Josh},
	date = {2019-12-13},
}

@article{spearing_ranking_2021,
	title = {Ranking, and other properties, of elite swimmers using extreme value theory},
	volume = {184},
	issn = {1467-985X},
	pages = {368--395},
	number = {1},
	journaltitle = {Journal of the Royal Statistical Society: Series A (Statistics in Society)},
	author = {Spearing, Harry and Tawn, Jonathan and Irons, David and Paulden, Tim and Bennett, Grace},
	date = {2021},
}

@article{guinness_observational_2020,
	title = {An Observational Study of the Effect of Nike Vaporfly Shoes on Marathon Performance},
	journaltitle = {{arXiv}:2002.06105 [stat]},
	author = {Guinness, Joseph and Bhattacharya, Debasmita and Chen, Jenny and Chen, Max and Loh, Angela},
	urldate = {2020-04-25},
	date = {2020-02-14},
}

@article{einmahl_limits_2019,
	title = {Limits to Human Life Span Through Extreme Value Theory},
	volume = {114},
	issn = {0162-1459},
	pages = {1075--1080},
	number = {527},
	journaltitle = {Journal of the American Statistical Association},
	author = {Einmahl, Jesson J. and Einmahl, John H. J. and Haan, Laurens de},
	date = {2019-07-03},
}

@article{katz_statistics_2002,
	title = {Statistics of extremes in hydrology},
	volume = {25},
	issn = {0309-1708},
	pages = {1287--1304},
	number = {8},
	journaltitle = {Advances in Water Resources},
	author = {Katz, Richard W and Parlange, Marc B and Naveau, Philippe},
	date = {2002-08-01},
}

@incollection{holmes_statistical_2008,
	title = {Statistical Analysis of Large Wildfires},
	isbn = {978-1-4020-4369-7},
	pages = {59--77},
	booktitle = {The Economics of Forest Disturbances: Wildfires, Storms, and Invasive Species},
	author = {Holmes, Thomas and Huggett, Robert and Westerling, A.},
	publisher = {Springer, Dordrecht},
	date = {2008},
}

@article{tucker2017unlikeliness,
  title={The unlikeliness of an imminent sub-2-hour marathon: historical trends of the gender gap in running events},
  author={Tucker, Ross and Santos-Concejero, Jordan},
  journal={International journal of sports physiology and performance},
  volume={12},
  number={8},
  pages={1017--1022},
  year={2017},
  publisher={Human Kinetics, Inc.}
}

@article{hunter2015two,
  title={The two-hour marathon: what's the equivalent for women?},
  author={Hunter, Sandra K and Joyner, Michael J and Jones, Andrew M},
  journal={Journal of Applied Physiology},
  volume={118},
  number={10},
  pages={1321-1323},
  year={2015},
  publisher={American Physiological Society Bethesda, MD}
}

@article{joyner_two-hour_2011,
	title = {The two-hour marathon: who and when?},
	volume = {110},
	issn = {8750-7587},
	pages = {275--277},
	number = {1},
	journaltitle = {Journal of Applied Physiology},
	author = {Joyner, M. J. and Ruiz, J. R. and Lucia, A.},
	date = {2011-01-01}
}

@book{gumbel_statistics_1958,
	title = {Statistics of Extremes},
	isbn = {978-0-231-89131-8},
	publisher = {Columbia University Press},
	author = {Gumbel, E. J.},
	date = {1958},
}

@article{bermon2021effect,
  title={Effect of advanced shoe technology on the evolution of road race times in male and female elite runners},
  author={Bermon, St{\'e}phane and Garrandes, Fr{\'e}d{\'e}ric and Szabo, Andras and Berkovics, Imre and Adami, Paolo Emilio},
  journal={Frontiers in Sports and Active Living},
  volume={3},
  pages={46},
  year={2021},
  publisher={Frontiers}
}

@article{senefeld2021technological,
  title={Technological advances in elite marathon performance},
  author={Senefeld, Jonathon W and Haischer, Michael H and Jones, Andrew M and Wiggins, Chad C and Beilfuss, Rachel and Joyner, Michael J and Hunter, Sandra K},
  journal={Journal of Applied Physiology},
  year={2021},
  publisher={American Physiological Society Rockville, MD}
}

@article{kipp2019extrapolating,
  title={Extrapolating metabolic savings in running: implications for performance predictions},
  author={Kipp, Shalaya and Kram, Rodger and Hoogkamer, Wouter},
  journal={Frontiers in physiology},
  volume={10},
  pages={79},
  year={2019},
  publisher={Frontiers}
}

\newpage
\begin{appendices}
\section{Theory and Model}\label{app:model}
\subsection{Extremes for identically distributed variables}
Extreme value theory (EVT) is a branch of statistics which studies the tails of probability distributions. It was first developed for block maxima (\citeauthor{gumbel_statistics_1958}, 1958) analysis, but the Peaks Over Threshold (POT) method (\citeauthor{davison_models_1990}, 1990) is often preferred, as it uses all the most extreme data, rather than just the maxima, typically leading to more efficient inference. Let $X$ be a random variable with distribution function $F$, if there exist random sequences $a_n$, $b_n > 0$ such that
\begin{equation}
    n\left\{(1-F(a_nx+b_n)\right\}\longrightarrow-\log G(x)
    \label{limiting}
\end{equation}
as $n\longrightarrow \infty$ is a non-degenerate limiting distribution, then for a large enough threshold $u$ we can use the approximation
\begin{align}
     Pr(X >x | X>u)\approx H_u(x)&=
    \begin{cases}
        1-\left[1+\xi\{(x-u)/\sigma_u\}\right]^{-1/\xi} & \xi\neq 0, \\
        1-\exp\{-(x-\mu)/\sigma_u\}, \quad & \xi=0,
    \end{cases}
   & x \in \R,
      \label{ett}
\end{align}

where $\sigma_u=\sigma+\xi(u-\mu)>0$, $a_+=max(a,0)$. If $\xi<0$ then $x$ must lie in the interval $[0,x_H]$, where $x_H=u-\sigma_u/\xi$ is the upper limit of the distribution, whereas if $\xi\geq 0$, $x$ can take any positive value. The limit distribution $H_u$, called Generalized Pareto distribution (GPD) motivates an approximation for large $u$, giving a model for the distribution of the exceedances above such threshold, regardless of the distribution $F$.

 Given a large enough sample of $n$ independent identically distributed (IID) observations, in the POT approach a threshold $u$ is carefully chosen, and exceedances can be used to estimate the parameters of the GPD. Threshold choice can be rather subjective and case-dependent, and is subject to a bias-variance trade-off. In this paper we base our choice on graphical diagnostics; however, other alternative methods might also be suitable; see \citeauthor{scarrott_review_2012} (2012) for a detailed review of these techniques.
 
It is remarkable that the rate of the frequency of exceedances above the threshold $u$ can be derived in a fashion that gives way to a more complete perspective of exceedances modelling, using point process models. Let $X_i$ be IID random variables with distribution function $F$, we define
 \begin{equation}
     N_n(x)=\sum_{i=1}^{n}\mathbbm{1}(X_i>a_{nx}+b_n),
 \end{equation}
 where $\mathbbm{1}(A)$ is an indicator whether the event $A$ occurs. It follows that $N_n(x)\sim Binomial(n,1-F(a_nx+b_n))$ with mean $n\left\{(1-F(a_nx+b_n)\right\}$, and using  the classical Poisson limit of the binomial distribution,
\begin{equation}
    N_n(x)\longrightarrow N(x)\sim Poisson(\lambda),
    \label{eq:poi}
\end{equation}
 where $\lambda=\{1+\xi(x-\mu)/\sigma\}_+^{-1/\xi}$.
 
 Therefore we can construct a model for extreme tails with two components: a model for the number of exceedances, given by (\ref{eq:poi}), which is Poisson distributed with mean $\lambda=\{1+\xi(x-\mu)/\sigma\}_+^{-1/\xi}$, and a model for the distribution of the exceedances, which is GPD distributed, following $H_u(x)$.
 
  Consider the sequence of point processes on $\mathbb{R}^2$ (\citeauthor*{coles_introduction_2001}, 2001)
 \begin{equation}
     P_n=\left\{\left(\frac{i}{n+1},\frac{X_i-b_n}{a_n}\right): i=1,\dots,n     \right\},
 \end{equation}
 where the scaling $1/(n+1)$ in the first coordinate ensures that the time axis is continuous on (0, 1), and the sequences $a_n$, $b_n$ are defined in (\ref{limiting}). More precisely, on regions of the form $[0,1]\times(u,\infty)$, where
 $u$ is large enough such that (\ref{ett}) approximately holds, we have
 have that $P_n\longrightarrow P$ as $n\longrightarrow\infty$, where $P$ is a non-homogeneous Poisson Process. 
 Consequently, the integrated measure $\Lambda$ of $P$ on $\mathcal{A}_{1,u}=[0,1]\times(u,\infty)$ is given by
 \begin{equation}
     \Lambda(\mathcal{A}_{1,u})=\left\{1+\xi\left(\frac{u-\mu}{\sigma}\right)\right\}_+^{-1/\xi},
 \end{equation}
and its intensity function is
\begin{equation}
    \lambda(t,x)=\frac{1}{\sigma}\left\{1+\xi\left(\frac{x-\mu}{\sigma}\right)\right\}_+^{-1/\xi-1}=\lambda(x),
    \end{equation}
with $x>u$ and $0<t\leq 1$. For statistical inference we assume that for large enough $n$, $P_n\sim P$ is a good approximation. The scaling coefficients $a_n,b_n$, can be absorbed into the intensity function, so we work directly with the series $\left\{\left(\frac{i}{n+1},X_i\right): i=1,\dots,n   \right\}$. Therefore, for a region of the form $\mathcal{A}_{1,u}=[0,1]\times(u,\infty)$, containing $n$ points $\left\{\mathbf{x}=(t_1,x_1),\dots,(t_n,x_n)\right\}$, the likelihood for the parameters $\theta=(\mu,\sigma,\xi)$ is
\begin{equation}
    L(\theta;\mathbf{x})=\exp\left\{-\Lambda(A_{1,u})\right\}\prod_{i=1}^n\lambda(x_i).
    \label{like}
\end{equation}

\subsection{Extremes of Non-Stationary sequences}
The extreme value models derived so far are built on the assumption of IID variables. However, in our work, non-stationarity data arise due to the improvement of racing conditions over time, and the potential Vaporfly shoes. Therefore, we relax the identically distributed assumption by introducing a time-dependent structure, while keeping independence assumption. Indeed, the time variation for parameters $\theta(t)=\{\mu(t),\sigma(t),\xi(t)\}$ will translate into a time-dependent rate of exceedances, and distribution of such exceedances. Under this covariate structure, the intensity of the non-homogeneous Poisson process $P$ will be
\begin{equation}
\lambda(t,x)=  \frac{1}{\sigma(t)}\left[1+\xi(t)\left\{\frac{x-\mu(t)}{\sigma(t)}\right\}\right]_+^{-1/\xi(t)-1}.
\label{lam}
\end{equation}
Now, in the general case where we have $n$ points $\left\{\mathbf{x}=(t_1,x_1),\dots,(t_n,x_n)\right\}$ in the region $\mathcal{A}_{T,u}=[0,T]\times(u,\infty)$, the integrated intensity becomes
\begin{equation}
 \Lambda(\mathcal{A}_{T,u})=\int_0^T\left[1+\xi(t)\left\{\frac{x-\mu(t)}{\sigma(t)}\right\}\right]_+^{-1/\xi(t)}dt  , 
 \label{Lam}
\end{equation}
and the full likelihood is
\begin{equation}
    L\{\theta(\mathbf{t});\mathbf{x}\}=\exp\left\{  \Lambda(\mathcal{A}_{T,u}) \right\}\prod_{i=1}^{n}\lambda(t_i,x_i).
    \label{full_lik}
\end{equation}
The parameters $\theta(\mathbf{t})=\left\{(\mu(t),\sigma(t),\xi(t)) \right\}$ are estimated by maximizing (\ref{full_lik}), and with such estimates, for a given time $t$, predictions about the number of exceedances can be made by integrating (\ref{lam}). The excess distribution at time $t$ will be given by
\begin{equation}
    Pr\left(X_{t}>x|X_{t}>u\right)=1-H_u(x,t)=\left[ 1+\xi(t)\left\{\frac{x-u}{\sigma_u(t)}\right\} \right]_+^{-\frac{1}{\xi(t)}},
    \label{gpd_ours}
\end{equation}
where $\sigma_u(t)=\sigma(t)+\xi(t)\{u-\mu(t)\}$.

\subsection{Model}
For most disciplines (and specially for marathon-men) a linear dependence on time for the scale parameter of the GP distribution of the exceedances was best suited in AIC terms. The following parametrisation was used to incorporate such structural time dependence.
\begin{align}
    \xi^{(d)}(t)&=\xi\\
    \mu^{(d)}(t)&=\mu_0^{(d)}+\beta^{(d)}y(t)+\gamma^{(d)}\mathbbm{1}_{\{y(t)\geq 2018\}}\\
    \sigma^{(d)}(t)&=\sigma_0^{(d)}+\xi^{(d)}\beta^{(d)}y(t)+\xi^{(d)}\gamma^{(d)}\mathbbm{1}_{\{y(t)\geq 2018\}}+\delta y(t)
\end{align}
where $d\in D$ is the superscript denoting discipline $d$, $y(t)$ is the year corresponding to time $t$, $\xi^{(d)},\mu_0^{(d)} \in \mathbb{R}$, $\sigma_0^{(d)}\in\mathbb{R}^+$ are the shape, location, and scale parameter of the Poisson process, $\beta \in \mathbb{R}$ controls the linear trend in $\sigma^{(d)}(t)$ and $\mu^{(d)}(t)$, $\gamma^{(d)}\in \mathbb{R}$ represents Vaporfly shoes effect, $\mathbbm{1}$ is the indicator function, and $2018$ is the year when the shoes started to be widely used in official races. Note that this parametrisation enforces the GPD scale parameter for exceedances above $u_d$ to change linearly with time.
\begin{align}
    \sigma_u^{(d)}(t)&=\sigma^{(d)}(t)+\xi^{(d)}\left\{u_d-\mu^{(d)}(t)\right\}\nonumber\\
    &=\sigma_0^{(d)}+\xi^{(d)}(u_d-\mu_0^{(d)}) +\delta y(t)\nonumber\\
    &:=\sigma_u^{(d)}+ \delta y(t)
\end{align}

\subsection{Expected running times of next new world record}
As derived in \citeauthor{spearing_ranking_2021} (2021), the expected new world record time for discipline $d$ at year $y$ will be
\begin{equation}
\mathbb{E}\left[X_y^{*(d)}\right] =\int_{r_d}^{x_{H,e}}x\frac{dH_{r_d}^{(d)}(x,y)}{dx}dx=r_d+\frac{\sigma_{r_d}^{(d)}(y)}{1-\xi},\quad \text{if }\xi<1,
\end{equation}
where $\sigma_{r_d}^{(d)}(y)=\sigma_0^{(d)}+\xi\left( r_d-\mu_0^{(d)}\right)+\delta y(t)$, $X_y^{*(d)}$ is the random variable denoting the running-time of a new world record for discipline $e$, set in year $y$, and $r_d$ is the current ($2019$) world record of discipline $d$, so that 
$r_d=max(\mathbf{X^{(d)}})$, with $\mathbf{X^{(d)}}$ the set of all observations for discipline $d$.

\subsection{Probability of breaking a world record in a given year}
Let $N_y^{(d)}$ be the number of exceedances of the threshold $u_d$ for discipline $d$ during year $y$, it is Poisson distributed with mean 
\begin{equation}
    \Lambda^{(d)}(\mathcal{A}_{y,u})=\left[1+\xi\left\{\frac{u_d-\mu^{(d)}(y)}{\sigma^{(d)}(y)}\right\}\right]_+^{-1/\xi}.
\end{equation}
Therefore, let $X^{(d)}_{1:N_y^{(d)}}=\left\{X_i^{(d)},i=1,\dots,N_y^{(d)}\right\}$, where $X_i^{(d)}\overset{iid}{\sim} H_u^{(d)}(y)$,
if we denote by $\text{Pr}(R^{(d)}_y)$ the probability that a world record for discipline $d$ is set in year $y$,
\begin{align}
\text{Pr}(R^{(d)}_y)=1-\exp\left\{-\Lambda^{(d)}(\mathcal{A}_{y,u})\bar{H}_u^{(d)}(r_d,y)\right\},
   \label{p_year}
\end{align}
where $\bar{H}_u^{(d)}(r_d,y) := 1-H_u^{(d)}(r_d,y)$.

\subsection{Time until next world record is set }
Let $T^{(d)}$ be the random variable describing the waiting time until a new world record is set for an discipline $e$, if we define $t_y=y-2020$, the probability $F_T^{(d)}(t_y)=\text{Pr}(T^{(d)}<t_y)$ that a world record for discipline $e$ is set before some year $y$ is
\begin{align}
    F_T^{(d)}(t_y)=1-\exp\left\{-\sum_{k=2020}^{y-1}\Lambda^{(d)}(\mathcal{A}_{k,u})\bar{H}_u^{(d)}(r_d,k)\right\}.
    \label{cumul}
\end{align}

We can further estimate the expected waiting time until the world record is broken for any discipline $e$, which has the following expression
\begin{align}
    \mathbb{E}\left[T^{(d)}\right]=\text{Pr}(R_{2020})+\sum_{t=2}^\infty\left[ \text{Pr}(R_{2019+t})\sum_{k=1}^{t-1}\left\{1-\text{Pr}(R_{2019+k})\right\}\right],
\end{align}
where Pr$(R_y)$ is the probability that the world record is broken at year $y$, as described in (\ref{p_year}).

\subsection{Adjusting for AFT effect}
Let $x>u$ be a running-time recorded during year $y>2018$, when Vaporfly and other shoes with technological advances are widely used in official races. We denote by $x_c$ the \textit{corrected} or \textit{equivalent} time of $x$ if such shoes were not used. Its expression can be derived as in \citeauthor{spearing_ranking_2021} (2021), obtaining
\begin{equation}
    x_c=u_d+\frac{\sigma_{C,u}^{(d)}(y)}{\xi}\left\{\frac{\Lambda^{(d)}(\mathcal{A}_{y,u})\bar{H}_u^{(d)}(x,y)}{\Lambda^{(d)}_{C}(\mathcal{A}_{y,u})}-1\right\},
    \label{adj}
\end{equation}
where $\Lambda^{(d)}_{C}(\mathcal{A}_{y,u})$ has the form of $\Lambda^{(d)}(\mathcal{A}_{y,u})$ but with the corrected parameters
\begin{align}
    \mu_C^{(d)}(y)&=\mu_0^{(d)}+\beta y,\\
    \sigma_C^{(d)}(y)&=\sigma_0^{(d)}+\xi\beta y+\delta y,\\
    \sigma_{C,u}^{(d)}(y)&=\sigma_C^{(d)}(y)+\xi\left\{(u_d-\mu_C^{(d)}(y)\right\}.
\end{align}

\subsection{Breaking the 2h marathon}
Let $\text{Pr}(B2=y)$ be the probability the two-hour marathon being broken in a given year $y$, it follows from (\ref{p_year}) that 
\begin{equation}
    \text{Pr}(B2=y)=1-\exp\left\{-\Lambda^{(marM)}(\mathcal{A}_{y,u})\bar{H}_u^{(mar M)}(2h,y)\right\},
    \label{p_sub2}
\end{equation}
where $2h:=-7200$ and \textit{marM} refers to the marathon-men discipline.
Additionally, we could compute the cumulative probability of achieving a sub two-hour marathon before year $y$, which follows from (\ref{cumul})
\begin{equation}
    \text{Pr}(B2<y)=1-\exp\left\{-\sum_{k=2020}^{y-1}\Lambda^{(marM)}(\mathcal{A}_{k,u})\bar{H}_u^{(marM)}(2h,k)\right\}.
\end{equation}

\section{Model estimates}\label{app:model_estimates}

\begin{table}[H]
\caption{Parameter estimates (with $95\%$ confidence intervals) for the model.}\vspace{0.2cm}
\begin{tabular}{cccc}
\multicolumn{1}{c|}{discipline}             & $\sigma^{(d)}_0$ & $\mu^{(d)}_0$    & $\beta^{(d)}$  \\ \midrule
\multicolumn{1}{c|}{Marathon-men}      & 30.63 (30.40,33.95) & -7591 (-7594,-7584) & 12.51 (12.43,12.61) \\
\multicolumn{1}{c|}{Marathon-women}    & 101.95 (101.81,105.62) & -8418 (-8420,-8403) & 7.83 (7.73,8.00) \\
\multicolumn{1}{c|}{Half marathon-men} & 16.15 (16.01,16.57) & -3577 (-3578,-3574) & 3.64 (3.61,3.68) \\
\multicolumn{1}{c|}{Half marathon-women}        & 39.17 (39.98,40.92) & -4112 (-4115,-4105) & 11.66(11.62,11.74) \\ 
\multicolumn{1}{c|}{10km-men}          & 10.64 (10.49,11.00) & -1647 (-1647,-1645) & 1.02 (0.99,1.06) \\
\multicolumn{1}{c|}{10km-women}        & 17.74 (17.58,18.23) & -1863 (-1865,-1860) & 2.12 (2.09,2.18) \\ 

                                       &       &       &       \\
\multicolumn{1}{c|}{}                  & $\delta^{(d)}$ &\multicolumn{1}{c|}{$\gamma^{(d)}$} & $\xi$    \\ \midrule
\multicolumn{1}{c|}{}                  & 3.77 (3.73,3.84) & \multicolumn{1}{c|}{12.60 (9.54,19.46)} & \multirow{5}{*}{-0.251 (-0.248,-0.255)} \\
\multicolumn{1}{c|}{}                  & 0.36 (0.33,0.50)  &\multicolumn{1}{c|}{ 51.42 (50.08,60.12)} &  \\
\multicolumn{1}{c|}{}                  & 0.85 (0.83,0.88)  & \multicolumn{1}{c|}{0.70 (-2.39,2.86)}  &  \\
\multicolumn{1}{c|}{}                  & 2.59 (2.57,2.63)  &\multicolumn{1}{c|}{ 14.64(13.09,18.34)} &  \\
\multicolumn{1}{c|}{}                  & 0.29 (0.28,0.32)  & \multicolumn{1}{c|}{6.44 (5.79,8.29)} &  \\
\multicolumn{1}{c|}{}                  & 0.63 (0.61,0.67)  & \multicolumn{1}{c|}{13.74 (11.83,15.54)} & \\
\end{tabular}
\label{M2_estimates}
\end{table}

\begin{table}[H]
\caption{World records as in $2019$ and AFT-corrected times for records recorded before 2018.}\vspace{0.2cm}
\begin{tabular}{c|cc}
discipline             & World record 2019 & AFT-corrected world record 2019    \\ \midrule
Marathon-men      & 02:01:39       & 02:01:48 (-3s,+8s)  \\
Marathon-women    & 02:14:04        & 02:14:17 (-2s,+12s) \\
Half marathon-men & 00:58:01        & -  \\
Half marathon-women & 01:04:51        & 01:05:08 (-3s,+5s) \\
10km-men          & 00:26:38       & -  \\
10km-women        & 00:29:43        & -\\
\end{tabular}
\label{tb:corrected}
\end{table}

\section{Model checking}\label{app: plots}

\begin{figure}[H]
\centering
    \subfigure{\includegraphics[width=0.47\textwidth]{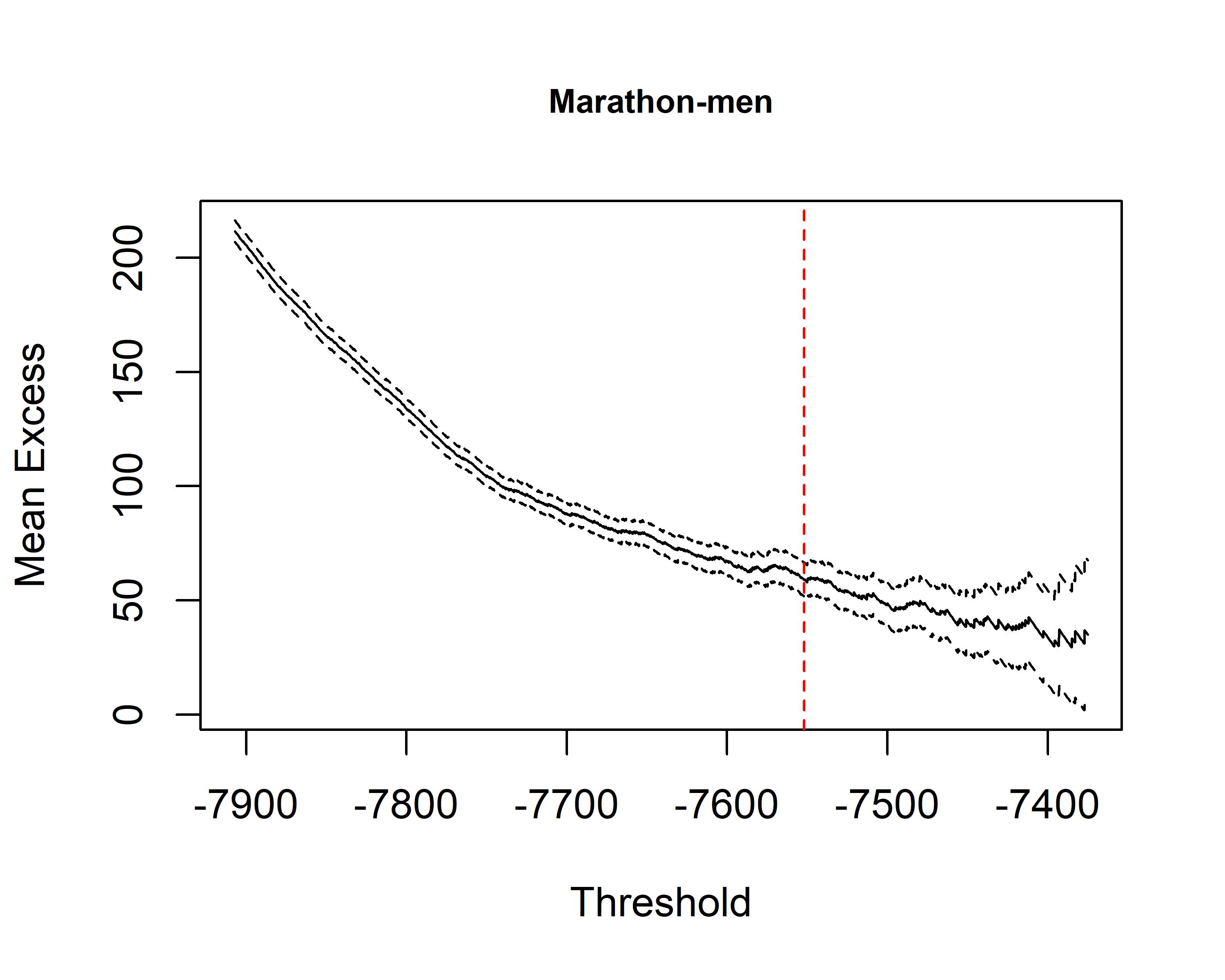}}
    \subfigure{\includegraphics[width=0.47\textwidth]{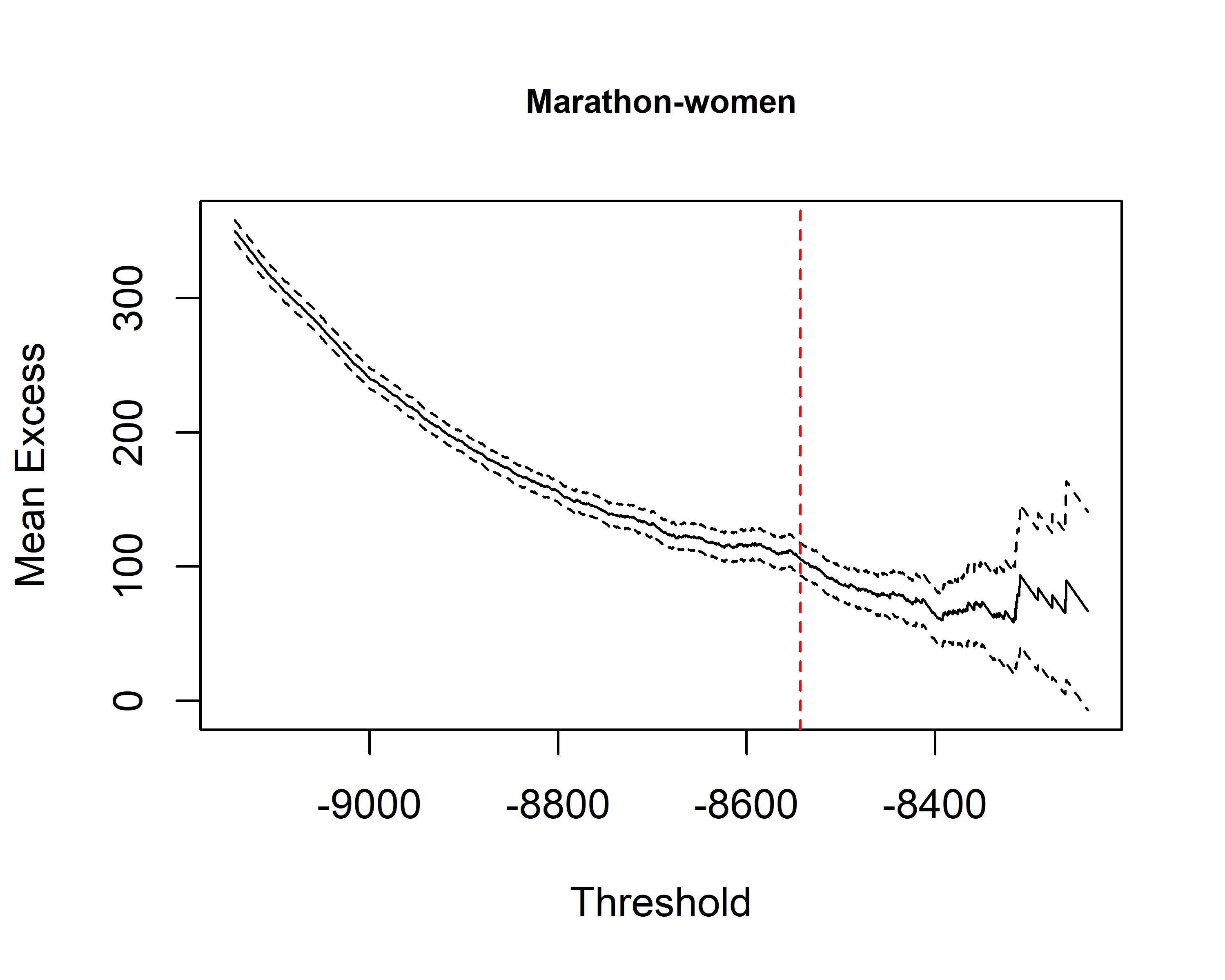}}
    \caption{Mean residual life plots, with $95\%$ confidence intervals. The red dashed line indicates the threshold used in our analysis.}
\end{figure}

\begin{figure}[H]
\centering
    \subfigure{\includegraphics[width=0.47\textwidth]{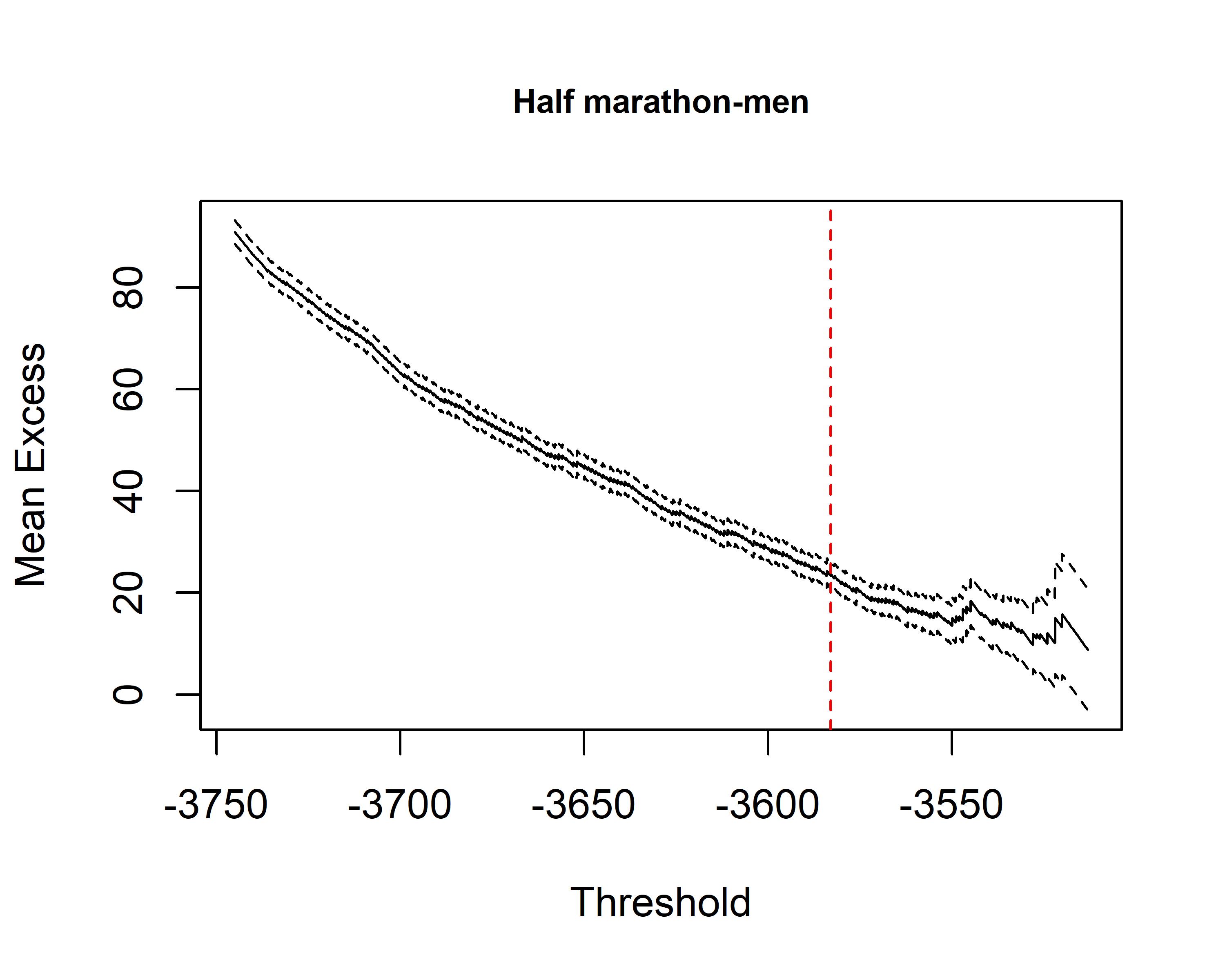}}
    \subfigure{\includegraphics[width=0.47\textwidth]{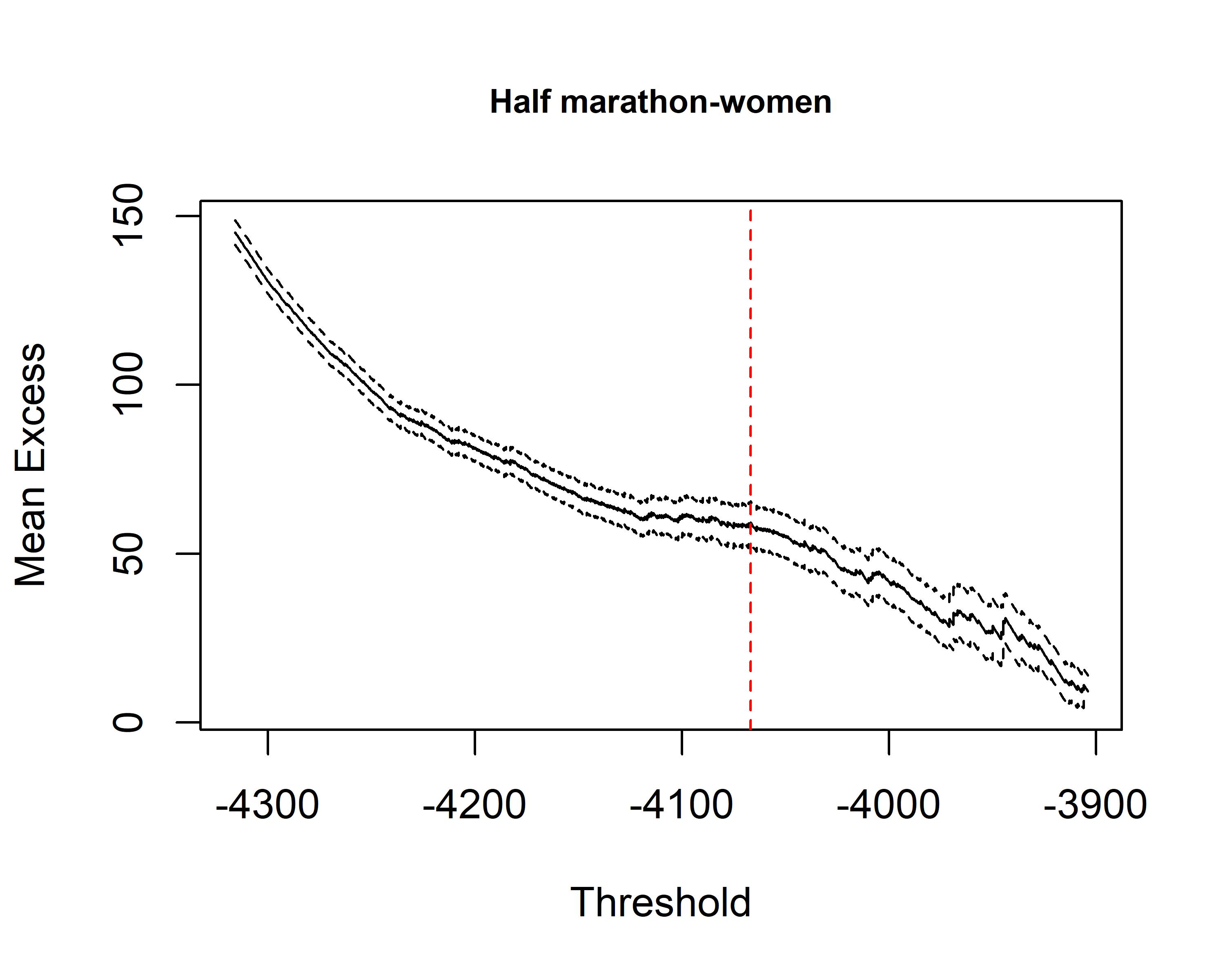}}
    \caption{Mean residual life plots, with $95\%$ confidence intervals. The red dashed line indicates the threshold used in our analysis.}
\end{figure}
\begin{figure}[H]
\centering
    \subfigure{\includegraphics[width=0.47\textwidth]{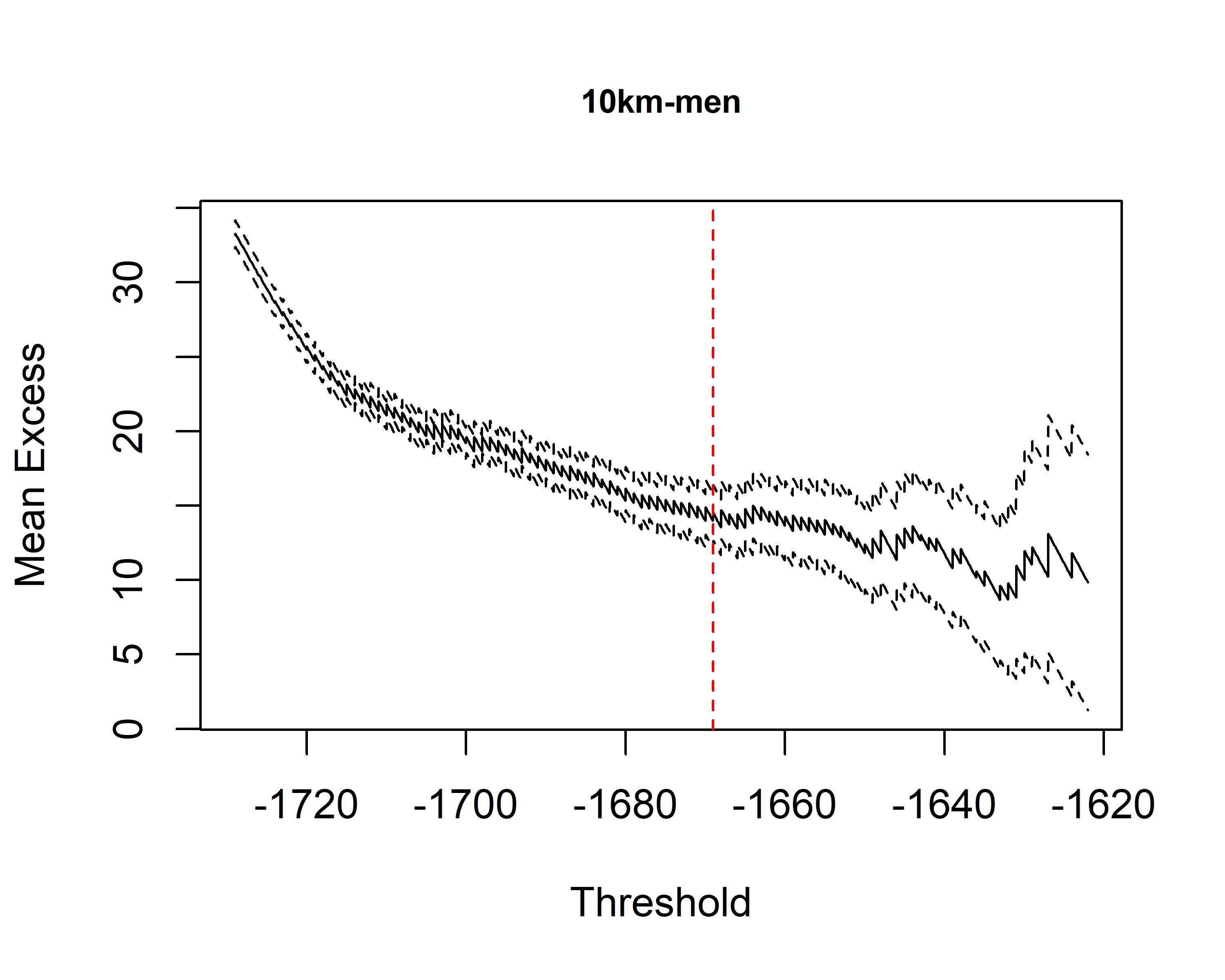}}
    \subfigure{\includegraphics[width=0.47\textwidth]{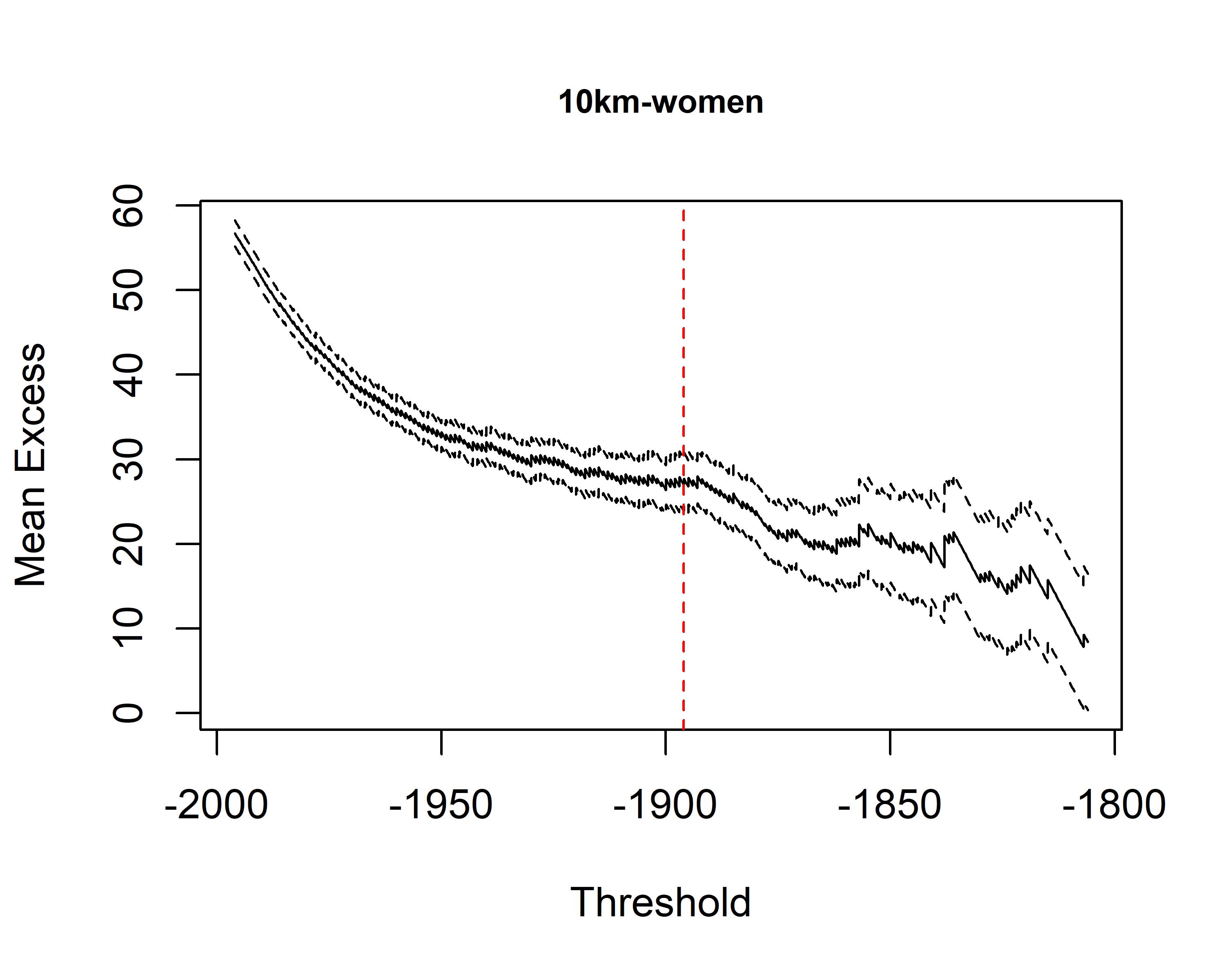}}

    \caption{Mean residual life plots, with $95\%$ confidence intervals. The red dashed line indicates the threshold used in our analysis.}
\end{figure}

\begin{figure}[H]
\centering
    \includegraphics[width=0.7\textwidth]{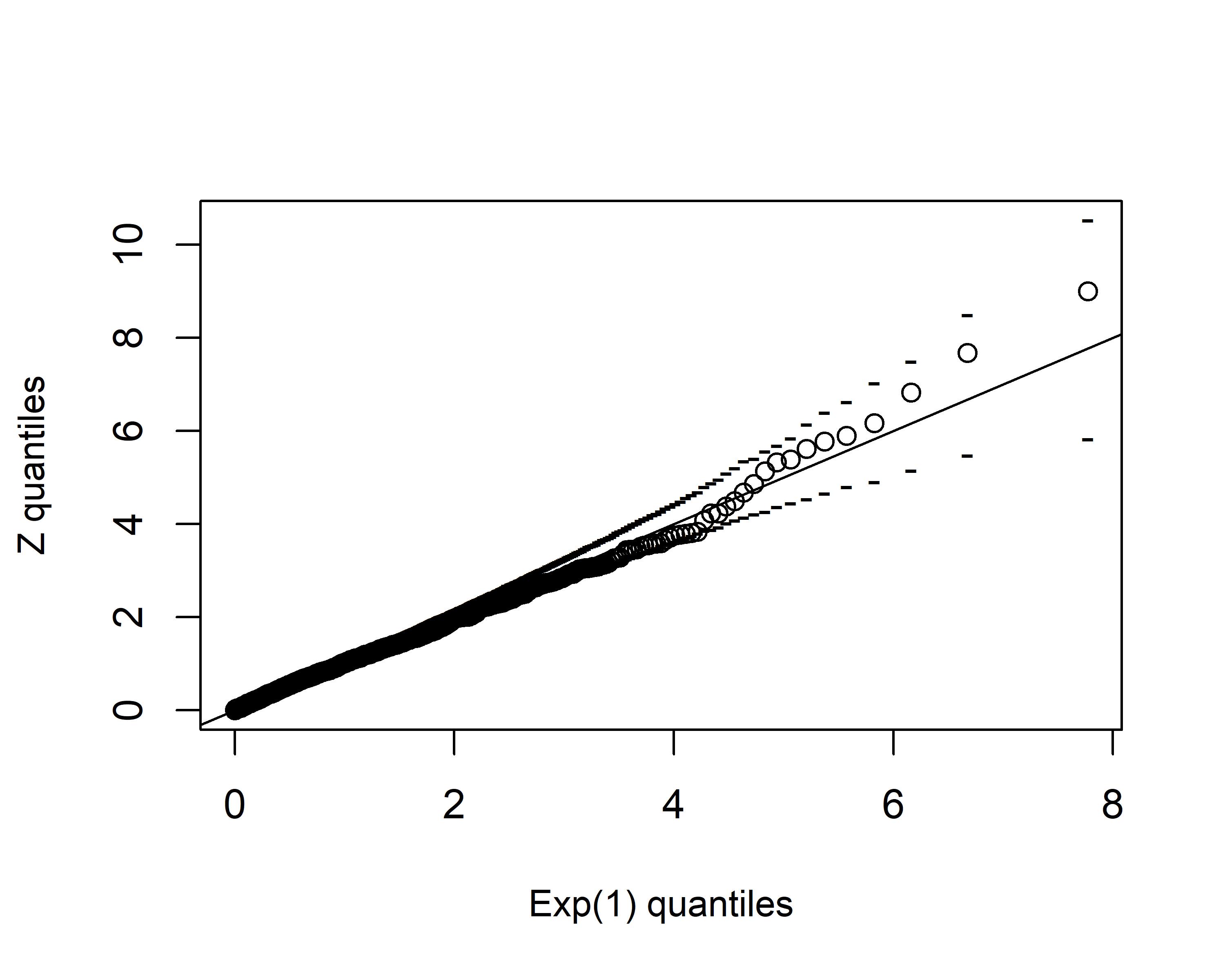}
    \caption{Diagnostic QQ plot for the model. The plot displays the log of the quantiles of the transformed observations for all disciplines, against the quantiles of a unit exponential distribution, with 95\% confidence intervals. }
    \label{qq_fit}
\end{figure}
\begin{figure}[H]
\centering
    \subfigure{\includegraphics[width=0.47\textwidth]{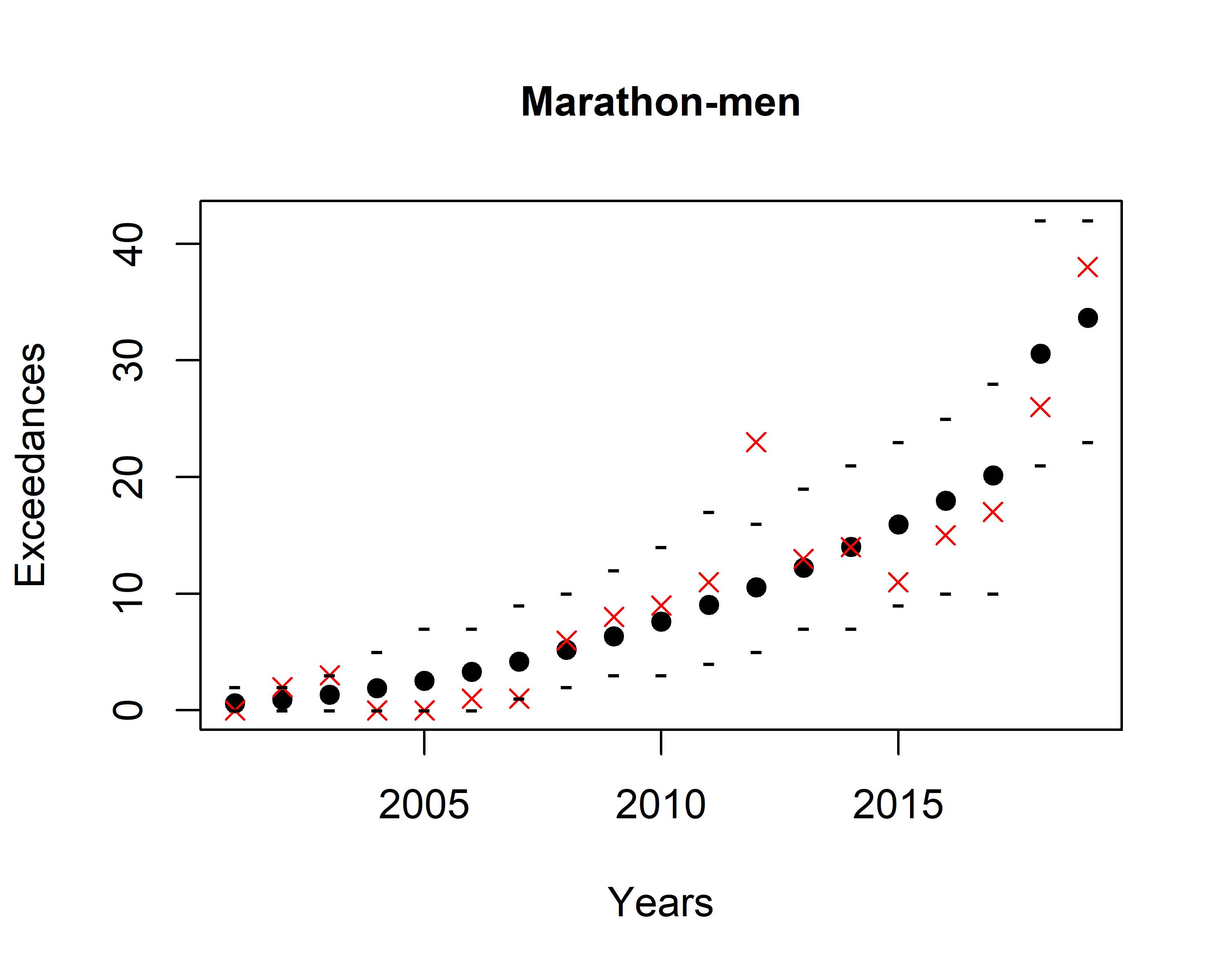}}
    \subfigure{\includegraphics[width=0.47\textwidth]{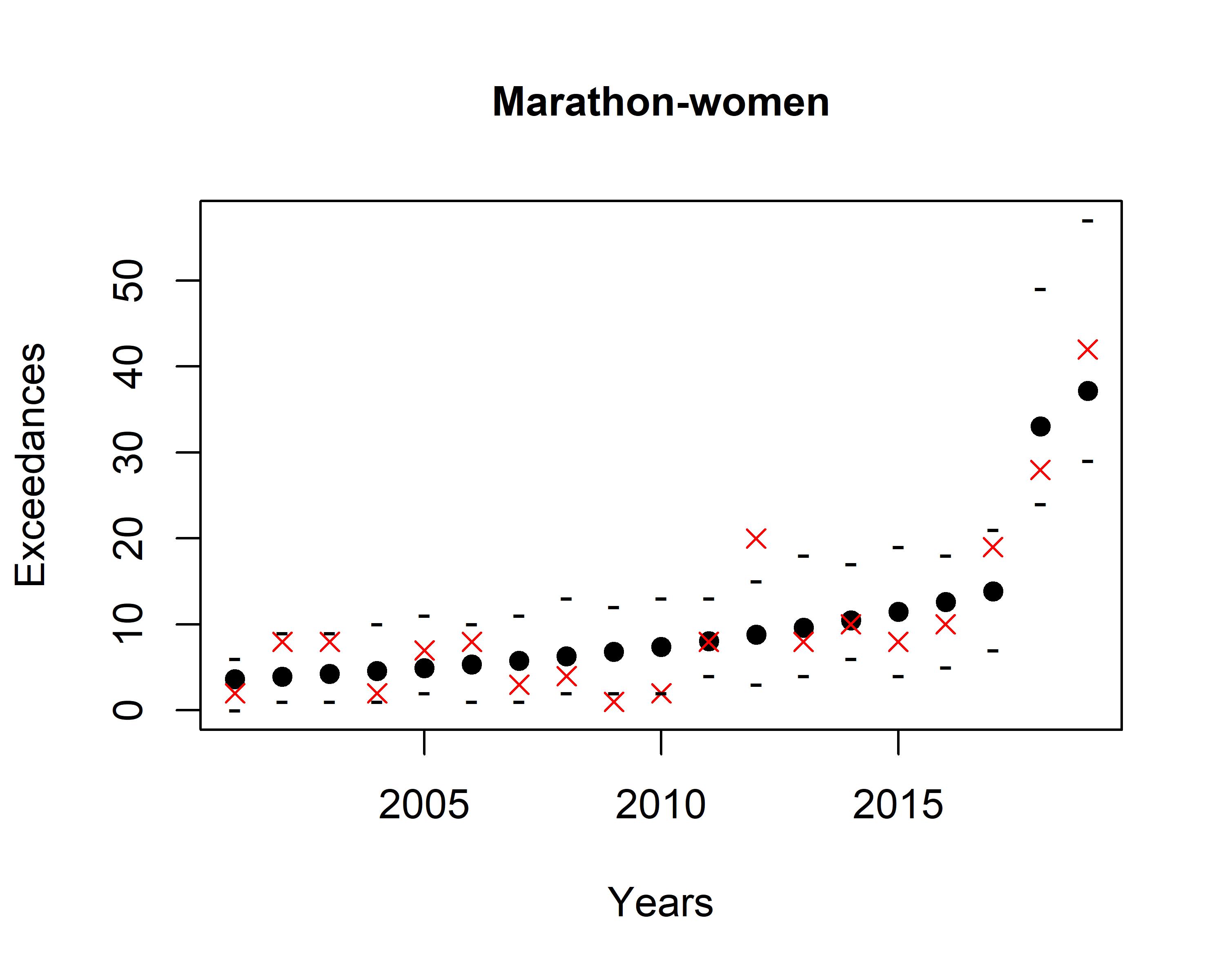}}
    \caption{Estimated expected (black circles) and observed (red crosses) exceedances above the threshold $u_d$ with $95\%$ confidence intervals (black dashes).}
\end{figure}

\begin{figure}[H]
\centering
    \subfigure{\includegraphics[width=0.47\textwidth]{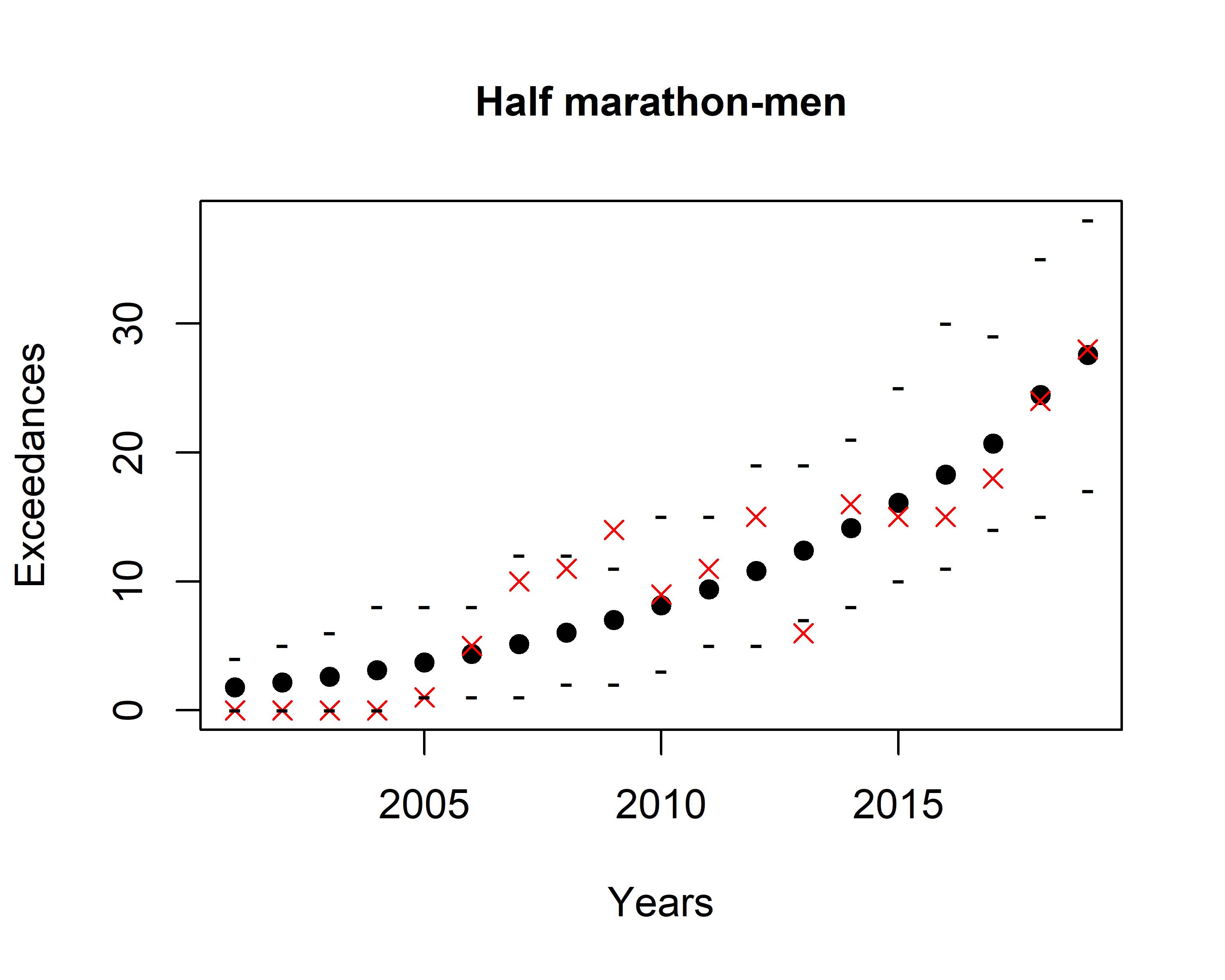}}
    \subfigure{\includegraphics[width=0.47\textwidth]{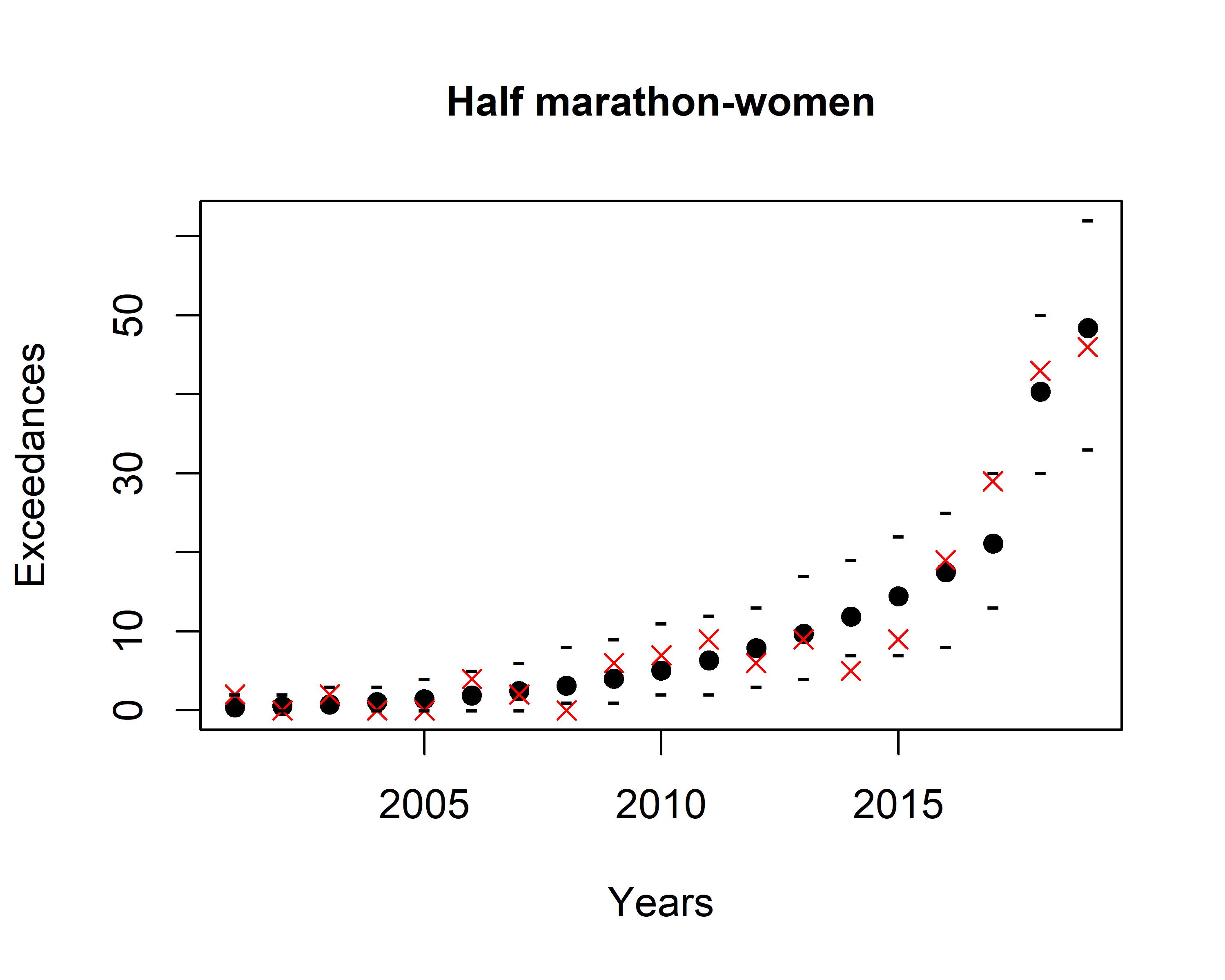}}
    \caption{Estimated expected (black circles) and observed (red crosses) exceedances above the threshold $u_d$ with $95\%$ confidence intervals (black dashes).}
\end{figure}
\begin{figure}[H]
\centering
    \subfigure{\includegraphics[width=0.47\textwidth]{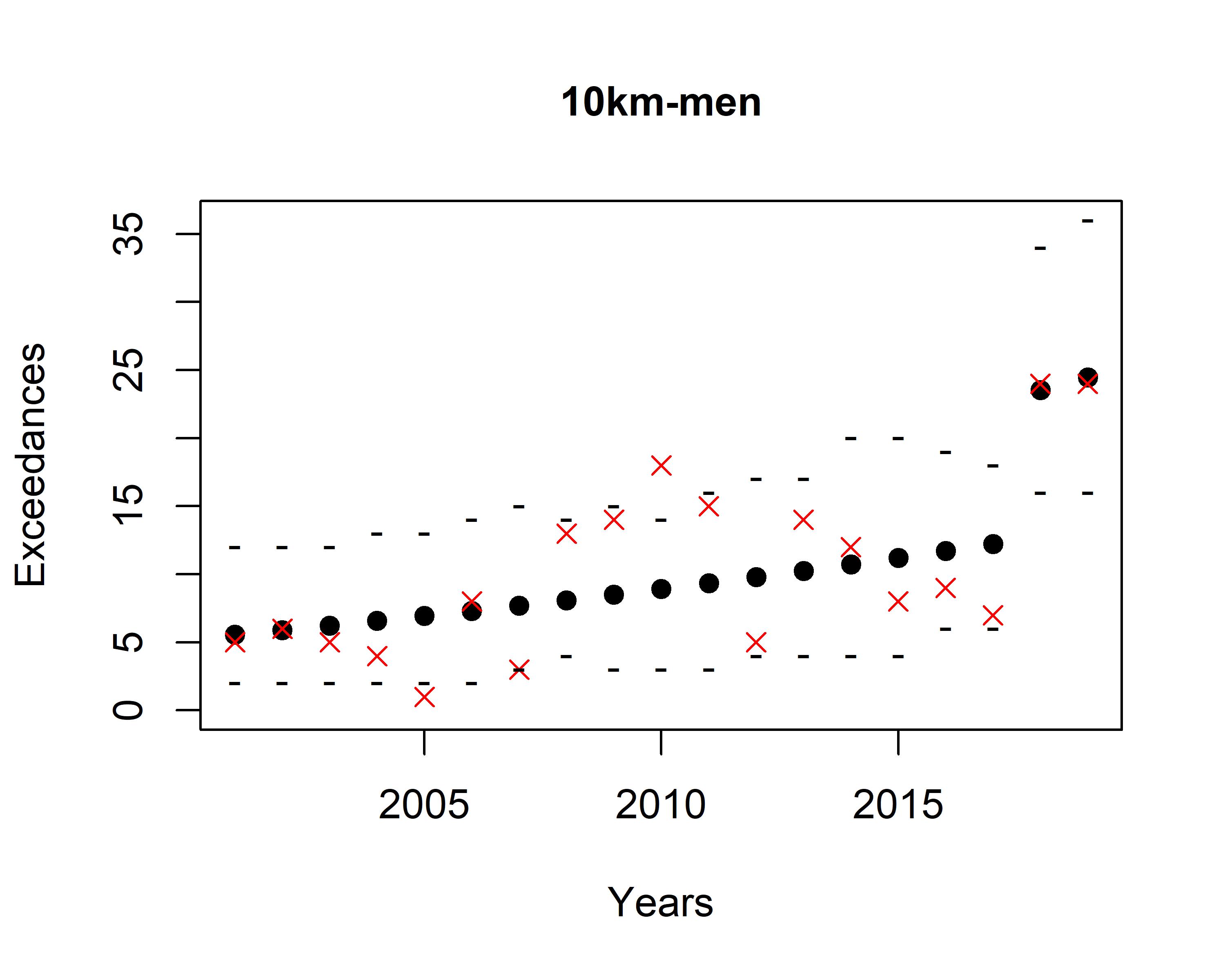}}
    \subfigure{\includegraphics[width=0.47\textwidth]{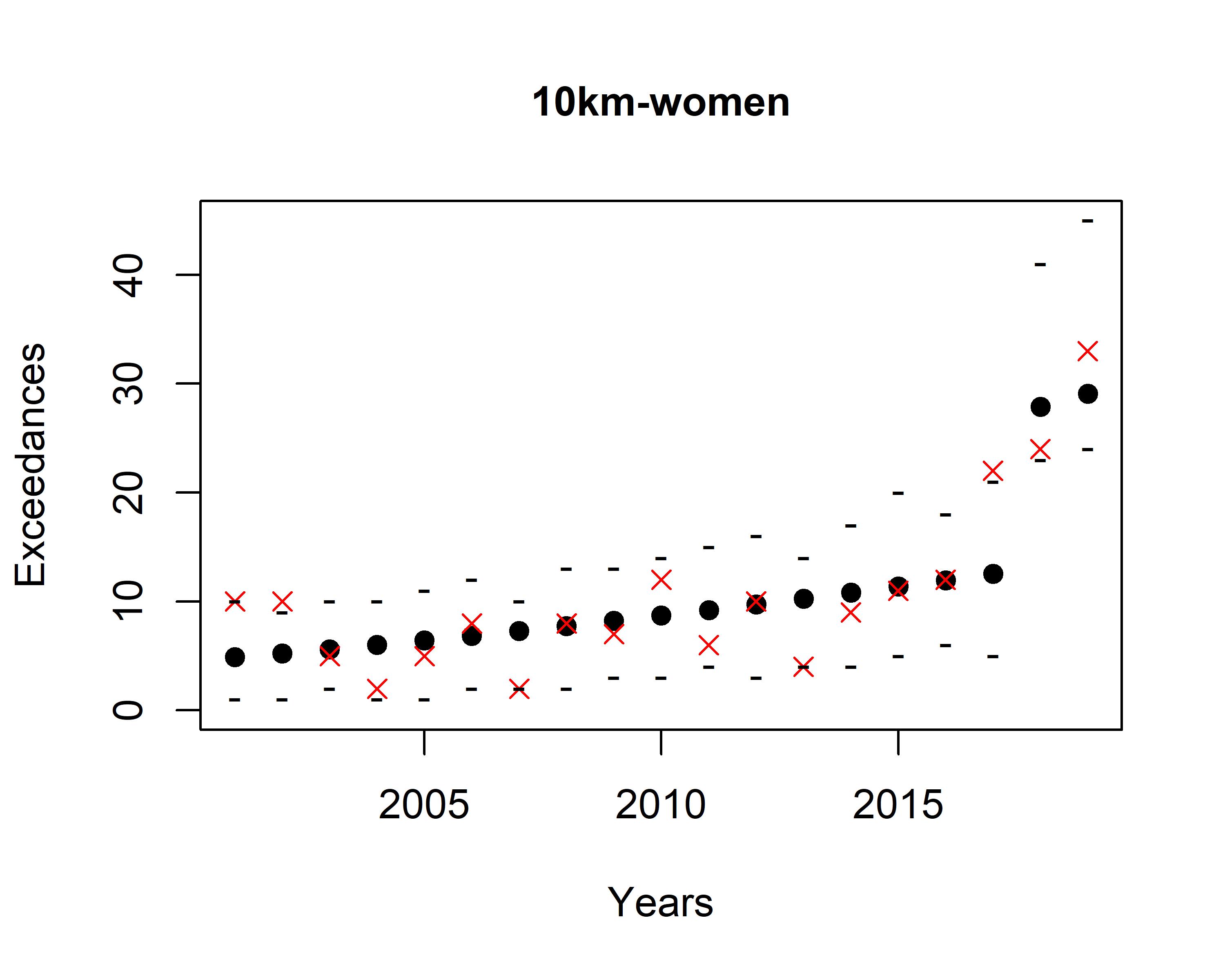}}

    \caption{Estimated expected (black circles) and observed (red crosses) exceedances above the threshold $u_d$ with $95\%$ confidence intervals (black dashes).}
\end{figure}

\end{appendices}

\end{document}